\documentclass[10pt]{iopart}
\usepackage{iopams}
\usepackage{psfrag}
\usepackage{pstricks}
\usepackage{array}
\usepackage{lscape}
\usepackage{graphics}
\usepackage{epsfig}

\renewcommand{\Re}{\mathrm{Re}}

\renewcommand{\vec}{\boldsymbol}

\newcommand{\op}[1]{\overline{\vec{#1}}}
\newcommand{\optilde}[1]{\widetilde{\vec{#1}}}

\newcommand{\ep}{\epsilon}

\newcommand{\p}[1]{\vec{p}_{#1}}

\newcommand{\hvec}[1]{\hat{\vec{#1}}}
\newcommand{\ehp}[1]{\hvec{e}_H(\p{#1})}
\newcommand{\evp}[2]{\hvec{e}_V^{#1}(\p{#2})}

\newcommand{\ez}{\hvec{e}_z}
\newcommand{\intr}[1]{\int  \rmd^3 \rr_{#1}\,}
\newcommand{\iintr}[2]{\int  \rmd^3 \rr_{#1}\, \rmd^3 \rr_{#2}\,}

\newcommand{\intp}[1]{\int \frac{\rmd^2 \vec{p}_{#1}}{(2\pi)^2}\,}
\newcommand{\intpV}[2]{\int_{\scriptstyle{#2}} \frac{\rmd^2 \vec{p}_{#1}}{(2\pi)^2}\,}
\newcommand{\iintpV}[3]{\int \!\!\! \int_{\scriptstyle{#3}}\frac{\rmd^2 \vec{p}_{#1}}{(2\pi)^2}\frac{\rmd^2 \vec{p}_{#2}}{(2\pi)^2}\,}

\newcommand{\intk}[1]{\int \frac{\rmd^3 \vec{k}_{#1}}{(2\pi)^3}\,}

\newcommand{\dirac}[1]{\delta(#1)}
\newcommand{\no}{\nonumber}
\newcommand{\x}{\vec{x}}

\newcommand{\rr}{\vec{r}}

\newcommand{\E}{\vec{E}}

\newcommand{\kk}{\vec{k}}

\newcommand{\alp}[2]{\alpha_{#1}(\vec{p}_{#2})}

\newcommand{\Grue}[3]{\op{G}^{#1}_{S}(\rr_{#2},\rr_{#3},\omega)}
\newcommand{\GVru}[3]{\op{G}^{#1}_{SV}(\rr_{#2},\rr_{#3},\omega)}

\newcommand{\RpS}[3]{\op{S}^{\,#1}\left(\p{#2}\right|\left.\p{#3}\right)}

\newcommand{\RS}[1]{\op{S}^{\,#1}}

\newlength{\Hlarger}
\begin{document}

\title[Electromagnetic wave scattering from a random layer with rough interfaces I]{Electromagnetic wave scattering from a random layer with rough interfaces I:  Coherent field}

\author{Antoine Soubret\dag \footnote[3]{asoubret@hms.harvard.edu} and G\'erard Berginc\ddag
 }

\address{\dag\ NOAA, Environmental Technology Laboratory, 325 Broadway, Boulder  CO 80305-3328}

\address{\ddag\ Thal\`es Optronique, Bo\^{\i}te Postale 55, 78233 Guyancourt Cedex, France}

\begin{abstract}
The problem of an electromagnetic wave scattered from a random medium layer with rough boundaries is formulated
using integral equations which involve two kinds of Green functions. The first one describes the wave scattered
by the random medium and the rough boundaries, and the second one which corresponds to the unperturbed Green
functions describes the scattering by an homogeneous layer with the rough boundaries. As these equations are
formally similar to classical equations used in scattering theory by an infinite random medium, we will be able
to apply standard procedures to calculate the coherent field. We will use the coherent potential approximation
where the correlations between the particles will be taken into account under the quasi-crystalline
approximation.
\end{abstract}

%Uncomment for PACS numbers title message
%\pacs{00.00, 20.00, 42.10}

% Uncomment for Submitted to journal title message
%\submitto{\JPA}

% Comment out if separate title page not required
\maketitle
\section{Introduction}
Many studies on electromagnetic waves scattered by a random medium layer with rough boundaries have been
reported in recent
years~\cite{Giov1,Giov2,Pak3,Lam2,Lam3,Ulaby3,Fung,Fung2,Shin,Mudaliar1,Mudaliar2,Mudaliar3,Furutsu1,Furutsu2,Calvo1}.
Rigorous numerical methods have been developed~\cite{Giov1,Giov2,Pak3} but are computationally intensive or
limited to 2D geometry. Most often, the radiative transfer theory is used for the volumetric scattering with the
Kirchoff or small-perturbation method for imposing the boundary
conditions~\cite{Lam2,Lam3,Ulaby3,Fung,Fung2,Shin,Kong,Kong2001-3}. This method is well suited to compute the
scattered intensity but is based on phenomenological considerations. Thus, analytical theory has been developed
in order to describe the coupling between the random medium and the rough boundaries.
Furutsu~\cite{Furutsu1,Furutsu2} formulates the rough surface scattering problem with Dyson and Bethe-Salpeter
equations which permit treating the random medium and the rough boundaries on the same footing. Unfortunately,
this approach is formal, and the relationship between the radiative transfer theory and the classical rough
surface scattering theories~\cite{Kong2001-3,Bass,Ogilvy,Beckmann,Voro} is not straightforward.
Mudaliar~\cite{Mudaliar1,Mudaliar2,Mudaliar3} uses integral equations where the rough boundaries are treated
under a perturbative development. He shows that the intensity verifies a "generalized" transport equation. If
this approach is more numerically tractable than Furutsu's, the expressions obtained are still involved. This is
due to the choice of perturbative development to describe the scattering by the rough surfaces. In this paper,
we show that we can obtain the general expression, whatever the choice of the scattering theory used at the
boundaries, in introducing the scattering operators of the rough surfaces~\cite{Voro}. Furthermore, in
separating the surface and the volume scattering contributions with the help of Green functions, we will be able
to use well developed  analytical theories of waves scattered by an infinite random
medium~\cite{Tsang1,Kong,Kong2001-3,Apresyan,Kuz,Bara1,Lag1,Sheng1,Sheng2}. In this paper, which is the first
part of a series of three papers, we investigate the coherent field scattered by the rough surfaces and the
random medium. The contribution of the random medium will be taken into account in introducing an effective
permittivity, which is calculated under the Quasi-Crystalline Coherent Potential Approximation
(QC-CPA)~\cite{Kong,Kong2001-3}. The contribution due to the rough surface will be given by the average of the
scattering operators~\cite{Voro}. The calculation of the incoherent fields will be the subject of the following
papers, where the derivation of the radiative transfer equation will be detailed, and the  particular case of
 strongly diffusing random media will be treated using a vectorial diffusion approximation for Rayleigh scatterers.

\section{Geometry of the problem and formulation}
\begin{figure}[htbp]
   \centering
      \psfrag{k0}{$\vec{k}_0$}
      \psfrag{k}{$\vec{k}$}
      \psfrag{t0}{$\theta_0$}
      \psfrag{t1}{$\theta$}
      \psfrag{x}{$x$}
      \psfrag{z}{$z$}
       \psfrag{I}{}
      \psfrag{milieu}{{\small Random medium}}
      \psfrag{surface}{{\small Rough boundaries}}
      \psfrag{H}{$H$}
      \psfrag{m0}{$V_0,\,\ep_0$}
      \psfrag{m1}{$V_1,\,\ep_1 $}
      \psfrag{m2}{$V_2,\,\ep_2$}
      \psfrag{es}{$\ep_s$}
      \psfrag{z1}{$z=h_1(\x)$}
      \psfrag{z2}{$z=-H+h_2(\x)$}
         \epsfig{file=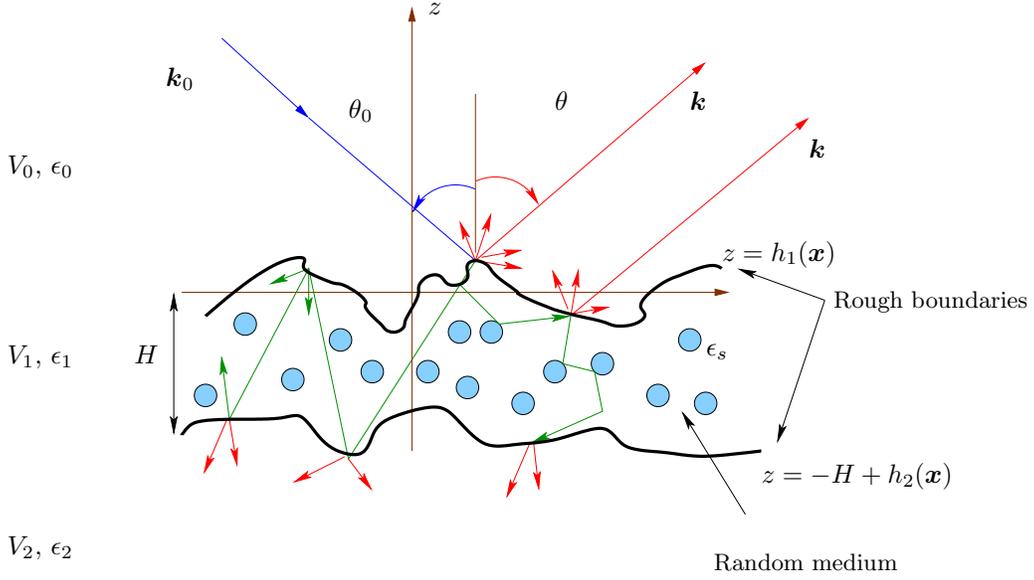,height=8cm}
      \caption{\label{Fig1}Random medium with rough boundaries.}
      \end{figure}

The geometry of the problem is shown in Figure \ref{Fig1}. Volumes $V_0$ and $V_2$ are homogeneous media with
permittivity $\ep_0(\omega)$ and $\ep_2(\omega)$. For simplicity we suppose that 
$\ep_0(\omega)$ is a real positive number. The random medium $V_1$ is made of spherical scatterers of
permittivity $\ep_s(\omega)$ in a background medium of permittivity $\ep_1(\omega)$. The boundaries are
described by the random functions $z=h_1(\vec{x})$ and $z=-H+h_2(\vec{x})$.

In the following, we consider  harmonic waves with $\e^{-\rmi\,\omega \,t}$ dependence. For a  point source
located at $\rr_0=\vec{x}_0+z_0\,\ez$ in the medium $V_0$, the field scattered by the rough surfaces and the
random medium at the point $\rr=\vec{x}+z\,\ez$ in the media $V_0$, $V_1$, $V_2$ are, respectively, given by the
dyadic Green functions~\cite{Kong,Tai} $\GVru{00}{}{0}$, $\GVru{10}{}{0}$, $\GVru{20}{}{0}$. Here, for
$\GVru{aa_0}{}{0}$, the upperscripts $a$,$a_0$ are, respectively, the receiver location and  the source
location. These Green functions satisfy~\cite{Tai,KongE}:
\begin{itemize}
\item Propagation equations :
\begin{eqnarray}
\fl\nabla\times\nabla\times\GVru{00}{}{0}-\ep_0(\omega)\,K_{vac}^2\,\GVru{00}{}{0}=\delta(\rr-\rr_0)\,\op{I}\,\label{cond0.1},\\
\fl\nabla\times\nabla\times\GVru{10}{}{0}-\ep_V(\rr,\omega)\,K_{vac}^2\,\GVru{10}{}{0}=0\,,\label{cond0.2}\\
\fl\nabla\times\nabla\times\GVru{20}{}{0}-\ep_2(\omega)\,K_{vac}^2\,\GVru{20}{}{0}=0\,\label{cond0.3},
\end{eqnarray}
with $K_{vac}\equiv\omega/c_{vac}$ the vacuum wave number, and $c_{vac}$ the light speed in the vacuum. The
permittivity $\ep_V(\rr,\omega)$ inside the random medium $V_1$ is defined by :
\begin{equation}
\ep_V(\rr,\omega)=\ep_1(\omega)+\sum_{j=1}^N (\ep_s(\omega)-\ep_1(\omega))\,\Theta_s(\rr-\rr_j)\,,\label{eVrr}
\end{equation}
where $\rr_1,\dots,\rr_N$ are the center of the particles, and $\Theta_s$ describes the spherical particle shape
:
\begin{equation}
\Theta_s(\rr)=\left\{\begin{array}{cc}1 & \mbox{if}\quad ||\rr||<r_s\\
  0 & \mbox{if}\quad ||\rr||>r_s\end{array}\right.\,,\label{Thetad}
\end{equation}
with $r_s$ the particle radius.
 \item  Boundary conditions on the upper rough surface:
\begin{eqnarray}
\fl&&\hvec{n}_{s1}\cdot\ep_0(\omega)\,\GVru{00}{}{0}=\hvec{n}_{s1}\cdot\ep_1(\omega)\,\GVru{10}{}{0}, \label{Bound0.1}\\
\fl&&\hvec{n}_{s1}\times\GVru{00}{}{0}=\hvec{n}_{s1}\times\GVru{10}{}{0 }\, , \label{Bound0.2}\\
\fl&&\hvec{n}_{s1}\cdot\left[\nabla\times\GVru{00}{}{0}\right]=\hvec{n}_{s1}\cdot\left[\nabla\times
\GVru{10}{}{0} )\right]\,
,\label{Bound0.3}\\
\fl&&\hvec{n}_{s1}\times\left[\nabla\times\GVru{00}{}{0}\right]=\hvec{n}_{s1}\times\,\left[\nabla\times\GVru{10}{}{0}\right],\label{Bound0.4}
\end{eqnarray}
where $\rr=\vec{x}+h_1(\vec{x})\ez$, and $\hvec{n}_{s1}$ is the exterior normal to the rough  surface
$z=h_1(\x)$:
\begin{equation}
\hvec{n}_{s1}\equiv\frac{\ez-\nabla h_1(\x)}{(1+(\nabla
h_1(\x))^2)^{1/2}}\,.
\end{equation}
\item  Boundary conditions on the bottom rough surface :
\begin{eqnarray}
\fl&&\hvec{n}_{s2}\cdot\ep_1(\omega)\,\GVru{10}{}{0}=\hvec{n}_{s2}\cdot\ep_2(\omega)\,\GVru{20}{}{0}, \label{Bound0.5}\\
\fl&&\hvec{n}_{s2}\times\GVru{10}{}{0}=\hvec{n}_{s2}\times\GVru{20}{}{0}\, , \label{Bound0.6}\\
\fl&&\hvec{n}_{s2}\cdot\left[\nabla\times\GVru{10}{}{0}\right]=\hvec{n}_{s2}\cdot\left[\nabla\times
\GVru{20}{}{0} )\right]\,
,\label{Bound0.7}\\
\fl&&\hvec{n}_{s2}\times\left[\nabla\times\GVru{10}{}{0}\right]=\hvec{n}_{s2}\times
\, \left[\nabla\times\GVru{20}{}{0}\right],\label{Bound0.8}
\end{eqnarray}
where $\rr=\vec{x}+[-H+h_2(\vec{x})]\ez$, and $\hvec{n}_{s2}$ is the exterior normal to the rough  surface
$z=h_2(\x)$:
\begin{equation}
\hvec{n}_{s2}\equiv\frac{-\ez+\nabla h_2(\x)}{(1+(\nabla h_2(\x))^2)^{1/2}}\,.
\end{equation}
\item Radiative conditions at infinity in the media $V_0$ and
$V_2$.
\end{itemize}
We also use Green functions where the source is situated in the medium $V_1$. The fields in the medium $V_0$,
$V_1$, $V_2$ are given by the Green functions $\GVru{01}{}{0}$, $\GVru{11}{}{0}$, $\GVru{21}{}{0}$  which
verify:
\begin{itemize}
\item Propagation equations:
\begin{eqnarray}
\fl \nabla\times\nabla\times\op{G}_{SV}^{01}(\rr,\rr_0)-\ep_0(\omega)\,K_{vac}^2\,\op{G}_{SV}^{01}(\rr,\rr_0)=0\,\label{Cond1.1},\\
\fl\nabla\times\nabla\times\op{G}_{SV}^{11}(\rr,\rr_0)-\ep_V(\rr,\omega)\,K_{vac}^2\,\op{G}_{SV}^{11}(\rr,\rr_0)=\delta(\rr-\rr_0)\,\op{I}\,,\label{Cond1.2}\\
\fl\nabla\times\nabla\times\op{G}_{SV}^{21}(\rr,\rr_0)-\ep_2(\omega)\,
K_{vac}^2\,\op{G}_{SV}^{21}(\rr,\rr_0)=0\,\label{Cond1.3},
\end{eqnarray}
\item Boundary conditions on the upper rough surfaces:
\begin{eqnarray}
\fl&&\hvec{n}_{s1}\cdot\ep_1(\omega)\,\op{G}_{SV}^{11}(\rr,\rr_0)=\hvec{n}_{s1}\cdot\ep_0(\omega)\,\op{G}_{SV}^{01}(\rr,\rr_0)\,
,\label{Bound1.1}\\
\fl&&\hvec{n}_{s1}\times\op{G}_{SV}^{11}(\rr,\rr_0)=\hvec{n}_{s1}\times \op{G}_{SV}^{01}(\rr,\rr_0)\, , \label{Bound1.2}\\
\fl&&\hvec{n}_{s1}\cdot\left[\nabla\times\op{G}_{SV}^{11}(\rr,\rr_0)\right]=\hvec{n}_{s1}\cdot
\left[\nabla\times\op{G}_{SV}^{01}(\rr,\rr_0)\right]\,
,\label{Bound1.3}\\
\fl&&\hvec{n}_{s1}\times\left[\nabla\times\op{G}_{SV}^{11}(\rr,\rr_0)\right]=\hvec{n}_{s1}\times
\,
\left[\nabla\times\op{G}_{SV}^{01}(\rr,\rr_0)\right],\label{Bound1.4}
\end{eqnarray}
\item Boundary conditions on the lower rough surface:
\begin{eqnarray}
\fl&&\hvec{n}_{s2}\cdot\ep_1(\omega)\,\GVru{11}{}{0}=\hvec{n}_{s2}\cdot\ep_2(\omega)\,\GVru{21}{}{0}, \label{Bound1.5}\\
\fl&&\hvec{n}_{s2}\times\GVru{11}{}{0}=\hvec{n}_{s2}\times\GVru{21}{}{0}\, , \label{Bound1.6}\\
\fl&&\hvec{n}_{s2}\cdot\left[\nabla\times\GVru{11}{}{0}\right]=\hvec{n}_{s2}\cdot\left[\nabla\times
\GVru{21}{}{0} )\right]\,
,\label{Bound1.7}\\
\fl&&\hvec{n}_{s2}\times\left[\nabla\times\GVru{11}{}{0}\right]=\hvec{n}_{s2}\times
\, \left[\nabla\times\GVru{21}{}{0}\right],\label{Bound1.8}
\end{eqnarray}
\item Radiative conditions at infinity in the media 0 and 2.
\end{itemize}
In order to to separate the contribution from the rough surfaces and the random medium, we introduce the dyadic
Green functions $\Grue{00}{}{0}$, $\Grue{10}{}{0}$, $\Grue{20}{}{0}$, $\Grue{01}{}{0}$, $\Grue{11}{}{0}$,
$\Grue{21}{}{0}$ which describe the scattering  by  the layer with the rough boundaries but \emph{without} the
random medium. These functions verify similar propagation equations and  boundary conditions as the Green
functions $\GVru{aa_0}{}{0}$, where the permittivity $\ep_V(\rr,\omega)$ due to the random medium is replaced by
an effective permittivity $\ep_e(\omega)$ in  equations (\ref{cond0.2}, \ref{Cond1.2}) and the permittivity
$\ep_1(\omega)$ is replaced by $\ep_e(\omega)$ in equations (\ref{Bound0.1}, \ref{Bound0.5}, \ref{Bound1.1},
\ref{Bound1.5}). This effective permittivity will be determined  using the Coherent-Potential Approximation
(CPA) with the Quasi-Crystalline Approximation (QCA)~\cite{Kong,Kong2001-3,Soven,Gyorffy,Korringa}. We will show
in Section \ref{Chap3greenDiff}
 how to write these Green functions with the help of scattering operators, which are common tools in scattering
 theory by rough surfaces~\cite{Voro}.
\section{Integral equations}
\begin{figure}
      \centering
        \psfrag{z}{$z$}
        \psfrag{z1}{$z=h_1(\x)$}
     \psfrag{z2}{$z=-H+h_2(\x)$}
        \psfrag{h}{$h$}
     \psfrag{ep0}{$\ep_0$}
     \psfrag{ep1}{$\ep_1$}
     \psfrag{epe}{$\ep_e$}
        \psfrag{ep2}{$\ep_2$}
     \psfrag{r}{$\rr$}
      \psfrag{r0}{$\rr_0$}
        \epsfig{file=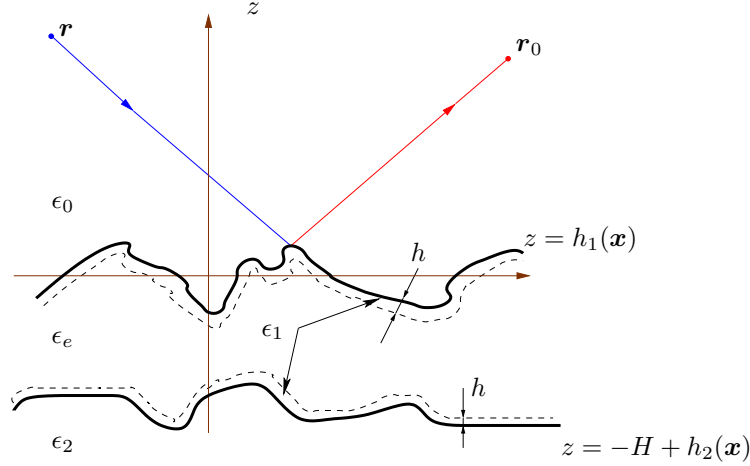,height=6cm}
\caption{\label{CoucheLim}Layers of thickness $h$ with a
permittivity $\ep_1$ around the boundaries .}
\end{figure}

The previous system of differential equations with boundary
conditions can be transformed into integral
equations~\cite{Kong,Kong2001-3,Tai}.
%equation in quantum scattering theory~\cite{Messiah}.
For a source in medium 0, we have
\begin{eqnarray}
\op{G}_{SV}^{00}&=&\op{G}_{S}^{00}+\op{G}_{S}^{01}\cdot\op{V}^{11}\cdot\op{G}_{S}^{10}+\op{G}_{S}^{01}\cdot\op{V}^{11}\cdot\op{G}_{SV}^{11}\cdot\op{V}^{11}\cdot\op{G}_{S}^{10}\,,\label{Lip1e}\\
\op{G}_{SV}^{10}&=&\op{G}_{S}^{10}+\op{G}_{S}^{11}\cdot\op{V}^{11}\cdot\op{G}_{S}^{10}+\op{G}_{S}^{11}\cdot\op{V}^{11}\cdot\op{G}_{SV}^{11}\cdot\op{V}^{11}\cdot\op{G}_{S}^{10}\,,\label{Lip2e}\\
\op{G}_{SV}^{20}&=&\op{G}_{S}^{20}+\op{G}_{S}^{21}\cdot\op{V}^{11}\cdot\op{G}_{S}^{10}+\op{G}_{S}^{21}\cdot\op{V}^{11}\cdot\op{G}_{SV}^{11}\cdot\op{V}^{11}\cdot\op{G}_{S}^{10}\,,\label{Lip3e}
\end{eqnarray}
and for a source in the medium 1
\begin{eqnarray}
\op{G}_{SV}^{01}&=&\op{G}_{S}^{01}+\op{G}_{S}^{01}\cdot\op{V}^{11}\cdot\op{G}_{SV}^{11}\,,\label{Lip4e}\\
\op{G}_{SV}^{11}&=&\op{G}_{S}^{11}+\op{G}_{S}^{11}\cdot\op{V}^{11}\cdot\op{G}_{SV}^{11}\,,\label{Lip5e}\\
\op{G}_{SV}^{21}&=&\op{G}_{S}^{21}+\op{G}_{S}^{21}\cdot\op{V}^{11}\cdot\op{G}_{SV}^{11}\,,\label{Lip6e}
\end{eqnarray}
with
\begin{eqnarray}
\op{V}^{11}(\rr,\rr_0,\omega)&=&\delta(\rr-\rr_0)\,\op{V}^1(\rr)\,,\label{Chap3Verr0}\\
\op{V}^1(\rr,\omega)&\equiv& K^2_{vac}(\ep_V(\rr,\omega)-\ep_e(\omega))\op{I}\,,\label{Chap3Ver}
\end{eqnarray}
and the following definition :
\begin{equation}
[\op{A}\cdot\op{B}](\rr,\rr_0)=\int_{V_1}
\rmd^3\,\rr_1\,\op{A}(\rr,\rr_1)\cdot\op{B}(\rr_1,\rr_0)\,.\label{Chap3Produitr}
\end{equation}
A direct demonstration of these equations involves integral theorems~\cite{Tsang6}, but it is easier to invoke
the uniqueness of the solution and verify \emph{a posteriori} that the integral equations
(\ref{Lip1e}-\ref{Lip6e}) satisfy the propagation equations and the boundary conditions. For example, to
demonstrate that equation \eref{Cond1.2} is verified, we apply the operator
$\nabla\times\nabla\times-\ep_e\,K^2_{vac}\op{I}$ on \eref{Lip5e}, and using the propagation equation satisfied
by $\op{G}_{S}^{11}$ with the definition in (\ref{Chap3Verr0}-\ref{Chap3Produitr}), we obtain
\begin{eqnarray}
\fl(\nabla\times\nabla\times- \ep_e\,K^2_{vac}\op{I})\cdot\op{G}_{SV}^{11}(\rr,\rr_0)&=(\nabla\times\nabla\times-\ep_e\,K^2_{vac}\op{I})\cdot\op{G}_{S}^{11}(\rr,\rr_0)\no\\
&\quad
+(\nabla\times\nabla\times-\ep_e\,K^2_{vac}\op{I})\cdot\op{G}_{S}^{11}\cdot\op{V}^{11}\cdot\op{G}_{SV}^{11}(\rr,\rr_0)\,,\no\\
&=\delta(\rr-\rr_0)\op{I}+\op{V}^{11}\cdot\op{G}_{SV}^{11}(\rr,\rr_0)\,,\no\\
&=\delta(\rr-\rr_0)\op{I}+K^2_{vac}(\ep_V(\rr)-\ep_e)\,\op{G}_{SV}^{11}(\rr,\rr_0)\,,\no
\end{eqnarray}
which is the propagation equation in \eref{Cond1.2}. By using the same procedure, we show that the propagation
equations (\ref{cond0.1}-\ref{cond0.3}, \ref{Cond1.1}-\ref{Cond1.3})  and the boundary conditions
(\ref{Bound0.2}-\ref{Bound0.4}, \ref{Bound0.6}-\ref{Bound0.8}) and (\ref{Bound1.2}-\ref{Bound1.4},
\ref{Bound1.6}-\ref{Bound1.8}) on the rough surfaces are satisfied. The boundary conditions at infinity in media
0 and 2 are specified by the choice of a retarded Green function for $\op{G}_{S}^{aa_0}$. However, due to the
introduction of the effective medium $\ep_e$, the boundary conditions (\ref{Bound0.1}, \ref{Bound0.5},
\ref{Bound1.1}, \ref{Bound1.5}) are not satisfied, and we obtain the following boundary conditions:
\begin{eqnarray}
&&\hvec{n}_{s1}\cdot\ep_e(\omega)\,\GVru{10}{}{0}=\hvec{n}_{s1}\cdot\ep_0(\omega)\,\GVru{00}{}{0}, \label{Bound3.1}\\
&&\hvec{n}_{s2}\cdot\ep_e(\omega)\,\GVru{10}{}{0}=\hvec{n}_{s2}\cdot\ep_2(\omega)\,\GVru{20}{}{0}, \label{Bound3.2}\\
&&\hvec{n}_{s1}\cdot\ep_e(\omega)\,\GVru{11}{}{0}=\hvec{n}_{s1}\cdot\ep_0(\omega)\,\GVru{01}{}{0}, \label{Bound3.3}\\
&&\hvec{n}_{s2}\cdot\ep_e(\omega)\,\GVru{11}{}{0}=\hvec{n}_{s2}\cdot\ep_2(\omega)\,\GVru{21}{}{0}.
\label{Bound3.4}
\end{eqnarray}
We see that in the left-hand side of equations (\ref{Bound3.1}-\ref{Bound3.4}), the permittivity is not
$\ep_1(\omega)$, as it must be, but is $\ep_e(\omega)$. If we had defined the Green function $\op{G}^{aa_0}_{S}$
describing the scattering by a homogeneous medium (with rough boundaries) with the permittivity $\ep_1$, the
problem would not exist. But if we want to use  the Coherent-Potential Approximation, we must introduce this
effective permittivity. We might go around this problem in changing the definition of the Green function
$\op{G}_{S}^{aa_0}$ where a small layer of arbitrary small thickness $h$ with permittivity $\ep_1$ is added
along the rough boundaries. (See Figure \eref{CoucheLim}.) These Green functions verify the boundary conditions
(\ref{Bound0.1}, \ref{Bound0.5}, \ref{Bound1.1}, \ref{Bound1.5}) where the permittivity $\ep_1$ is not replaced
by $\ep_e$ due to the added layers along the boundaries. Therefore, with this definition, equations
(\ref{Lip1e}-\ref{Lip6e}) verify the boundary conditions (\ref{Bound0.1}, \ref{Bound0.5}, \ref{Bound1.1},
\ref{Bound1.5}). However, the propagation equations (\ref{cond0.2}, \ref{Cond1.2}) are not satisfied since the
added layers produced  new contributions. But as we can choose the layers' thickness as thin as we want, we can
neglect the effect of these layers on the propagation equations.
In the following, we  won't take care of these boundary condition problems, and we will suppose that equations
(\ref{Lip1e}-\ref{Lip6e}) are solutions of our problem. The integral equations (\ref{Lip1e}-\ref{Lip6e}) are the
key point of our approach. We see that to calculate the field in  medium 0 or 2 (when the source is in  medium
0) with equations \eref{Lip1e} and \eref{Lip3e}, we first need  to determine the Green function
$\op{G}_{SV}^{11}$, where the source and the receiver are in  medium 1. This can be done with equation
\eref{Lip5e} where the only unknown is $\op{G}_{SV}^{11}$. If the permittivities of the medium 0, 1, and 2 and
the effective permittivity were equal ($\ep_e=\ep_0=\ep_1=\ep_2$), which means that scattering by the boundaries
does not take place, the Green function $\op{G}_{S}^{11}$ will be the Green function in an unbounded medium
(\ref{AppA}):
\begin{equation}
\op{G}^{\infty}_1(\rr,\rr_0)=\left(\op{I}+\frac{1}{K_e^2}\nabla\nabla\right)P.V.
\frac{\e^{\rmi\,K_e\,||\rr-\rr_0||}}{4\,\pi\,||\rr-\rr_0||}\,,
\end{equation}
where $K^2_e=\ep_e\,\omega^2/c^2_{vac}$, and equation \eref{Lip5e} becomes the usual equation used in scattering
theory by random media~\cite{Kong2001-3,Tsang6,Tsang3,Kuga2}. In taking into account the boundaries, we have to
change this  Green function for an infinite  random medium by Green functions taking into account the scattering
by the boundaries. However, it is worth mentioning  that the potential $\op{V}^{11}$ does not depend on the
boundaries, but only on the random medium. Because of this property, we will apply exactly the same procedures
developed in scattering theory by an infinite random medium, where the Green function for an unbounded medium
must be replaced by Green functions describing scattering by boundaries.
\section{Link between the Green functions $\op{G}_{S}^{11}$ and scattering operator}
\label{Chap3greenDiff} 
\begin{figure}[!hbtp]
   \begin{center}
      \psfrag{x}{$x$}
      %\psfrag{y}{$y$}
      \psfrag{z}{$z$}
      \psfrag{hx}{$z=h(\x)$}
      \psfrag{ehp0i}{$\ehp{0}$}
      \psfrag{evp00i}{$\evp{0-}{0}$}
      \psfrag{ehp}{$\ehp{}$}
      \psfrag{evp00}{$\evp{0+}{}$}
      \psfrag{evp01}{$\evp{1-}{}$}
      \psfrag{k0i}{$\vec{k}_{\vec{p}_0}^{0-}$}
      \psfrag{k0}{$\vec{k}_{\vec{p}}^{0+}$}
      \psfrag{k1}{$\vec{k}_{\vec{p}}^{1-}$}
      \psfrag{ep0}{$medium\,0 : \ep_0$}
      \psfrag{ep1}{$medium \,1 : \ep_e$}
      \psfrag{zmin}{$z_{min}=min_{\vec{x}}h(\x)$}
      \psfrag{zmax}{$z_{max}=max_{\vec{x}}h(\x)$}
      \epsfig{file=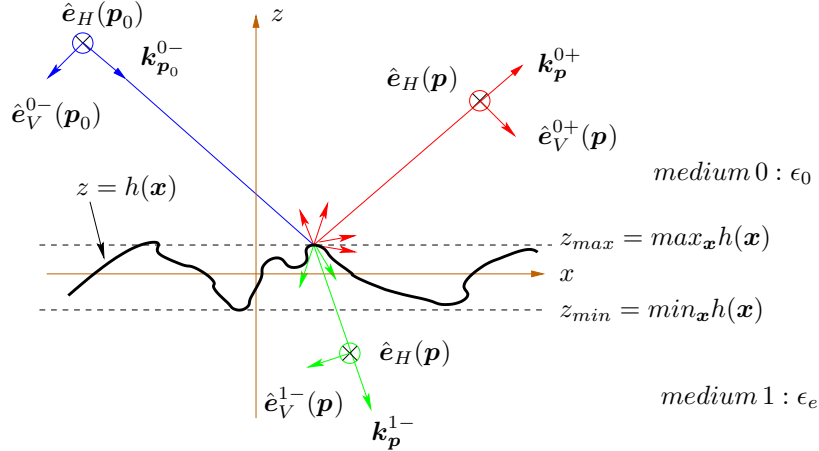,width=9cm}
\end{center}
        \caption{Incident plane wave with wave vector $\vec{k}_{\vec{p}_0}^{0-}$ reflected
        into medium 1 with the wave vector  $\vec{k}_{\vec{p}}^{0+}$ and transmitted into
        medium 1 with wave vector $\vec{k}_{\vec{p}}^{1-}$. The polarization basis are depicted.}
      \label{Fig3}
  \end{figure}

\begin{figure}[!htbp]
   \begin{center}
      \psfrag{kx}{$k_x$}
      \psfrag{ky}{$k_y$}
      \psfrag{kz}{$k_z$}
      \psfrag{phi}{$\phi$}
      \psfrag{t}{$\theta$}
      \psfrag{t0}{$\theta_0$}
      \psfrag{p}{$\p{}$}
      \psfrag{alp}{$\alp{0}{}$}
      \psfrag{alp0}{$\alp{0}{0}$}
      \psfrag{kp}{$\vec{k}_{\p{}}^{0+}$}
            \epsfig{file=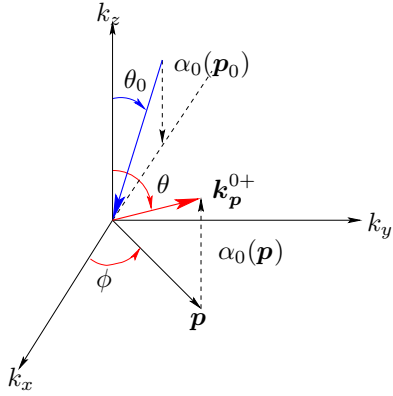,height=5cm}
            \caption{Wave vector decompositions}
%$\theta,\,\theta_0 \in [-\pi/2,\pi/2]$ et $\phi,\,\phi_0 \in [0,\pi[$,
      \label{Fig4}
   \end{center} \hfill
\end{figure}

In this section, we show how to express the Green functions $\op{G}_{S}^{11}$  with the
help of scattering operators which are common tools in scattering theory by rough surfaces~\cite{Voro}. These
operators describe the field scattered by a rough surface illuminated by an incident plane wave. (The Green
functions $\op{G}_{S}^{ab}$ describe the same phenomenon for a spherical incident wave.) For a rough surface
separating two semi-infinite homogeneous  media with permittivities $\ep_0$ and $\ep_e$ (Figure
\ref{Fig3}),
the wave equation can be simplified and transformed into the Helmholtz equation :
\begin{eqnarray}
(\Delta+K_0^2)\vec{E}^0(\rr)&=0,\;\;\; z>h(\x)\, ,\label{Hel1}\\
(\Delta+K_e^2)\vec{E}^1(\rr)&=0,\;\;\; z<h(\x)\, ,\label{Hel2}
\end{eqnarray}
with
\begin{equation}
K_0^2=\ep_0\,\left(\frac{\omega}{c_{vac}}\right)^2\,,\;\;\;
K_e^2=\ep_e\,\left(\frac{\omega}{c_{vac}}\right)^2\,,
\end{equation}
and transversality equation :
\begin{eqnarray}
\nabla\cdot\vec{E}^0(\x,z)&=0,\;\;\; z>h(\x)\, ,\label{divE0}\\
\nabla\cdot\vec{E}^0(\x,z)&=0,\;\;\; z<h(\x)\,.\label{divE1}
\end{eqnarray}
To find the fields $\vec{E}^0$ and $\vec{E}^1$, we need the
boundary conditions on the rough surface
$\rr_s=\vec{x}+h(\vec{x})\ez$:
\begin{eqnarray}
\hvec{n}_s\times
\vec{E}^0(\rr_s)&=&\hvec{n}_s\times \vec{E}^1(\rr_s)\, , \label{Pass1}\\
\hvec{n}_s\cdot\ep_0\,\vec{E}^0(\rr_s)&=&\hvec{n}_s\cdot\ep_e\,\vec{E}^1(\rr_s)\,
,\label{Pass2}\\
\hvec{n}_s\times\left[\nabla\times\vec{E}^0(\rr_s)\right]&=&\hvec{n}_s\times\left[\nabla\times\vec{E}^1(\rr_s)\right] \, ,\label{Pass3}\\
\hvec{n}_s\cdot\left[\nabla\times\vec{E}^0(\rr_s)\right]
&=&\hvec{n}_s\cdot\left[\nabla\times\vec{E}^1(\rr_s)\right] \, ,\label{Pass4}
\end{eqnarray}
and the radiation condition at infinity. For an incident plane
wave
$$\E^{0i}(\rr)=\E^{0i}(\p{0})\,\e^{i\,\p{0}\cdot\x-i\,\alp{0}{0}z}
$$  coming from  medium 0, the solution of these equations can
be written on the following form~\cite{Voro}:
\begin{eqnarray}
\fl\vec{E}^{0}(\x,z)=\vec{E}^{0i}(\x,z)+\vec{E}^{0s}(\x,z)\,,\\
\fl\vec{E}^{0s}(\x,z)=\intpV{}{}
  \op{R}^{10}(\p{}|\p{0})\cdot\E^{0\,i}(\p{0})\,\e^{i\,\vec{p}\cdot\x+i\,\alp{0}{}z}\;\;\mbox{for}\;\; z>max_{\vec{x}}\,h(\x)\no\,,\\
\label{E0bis}\\
\fl\vec{E}^{1}(\x,z)=\vec{E}^{1t}(\x,z)\,,\\
\fl\vec{E}^{1t}(\x,z)=\intpV{}{}
\,\op{T}^{10}(\p{}|\p{0})\cdot\E^{0\,i}(\p{0})\,\e^{i\,\vec{p}\cdot\x-i\,\alp{e}{}z}\;\;\mbox{for}\;\;z<min_{\vec{x}}\,h(\vec{x})
\, .\no\\ \label{E1bis}
\end{eqnarray}
Here~\footnote{In this definition, we need a precise the meaning of the square root because the integrand can be
negative or complex if the media are absorbing. Since the imaginary part of the permittivity is always positive
for an absorbing medium, we can use the following square root determination:
 \begin{equation}
\sqrt{z}=\left(\frac{|z|+\Re(z)}{2}\right)^{1/2}+\rmi\,\left(\frac{|
z|-\Re(z)}{2}\right)^{1/2}\,,
\end{equation} which corresponds to
classical square root operation for $z\in \mathbb{R}^+$,}
\begin{equation}
\alpha_0(\p{})=\sqrt{K_0^2-\p{}^2} \,,\;\;\; \alpha_e(\p{})=\sqrt{K_e^2-\p{}^2} \, .\label{alp}
\end{equation}
It can be easily checked that the propagation equations \eref{Hel1} and \eref{Hel2} are satisfied with the
representations (\ref{E0bis},\ref{E1bis}) and the definitions \eref{alp}. To satisfy the transversality
conditions \eref{divE0} and \eref{divE1}, we need to decompose the scattering operators
$\op{R}^{10}(\p{}|\p{0})$ and $\op{T}^{10}(\p{}|\p{0})$ on  an orthogonal basis perpendicular to propagation
vectors defined by $\p{}$. These vectors are given by the following formula in medium 0 and 1 (Figure
\ref{Fig4}):
\begin{eqnarray}
\vec{k}_{\p{}}^{0+}&= & \vec p+\alpha_0(\vec p)\ez \,,\label{transverse0}\\
\vec{k}_{\p{}}^{1-}&= & \vec p-\alpha_e(\vec p)\ez
\,.\label{transverse1}
\end{eqnarray}

\begin{figure}[htbp]
   \centering
      \psfrag{x}{$x$}
      \psfrag{ep0}{me0}
      \psfrag{R00}{$\scriptstyle \RpS{0+0-}{}{0}$}
      \psfrag{R1p0}{$\scriptstyle \RpS{1+\,0-}{}{0}$}
      \psfrag{R1mp0}{$\scriptstyle \RpS{1-\,0-}{}{0}$}
      \psfrag{R1p1p}{$\scriptstyle \RpS{1+\,1+}{}{0}$}
      \psfrag{incidente}{incident wave}
      \psfrag{m0}{$\ep_0$}
      \psfrag{m1}{$\ep_e$}
      \psfrag{m2}{$\ep_2$}
      \epsfig{file=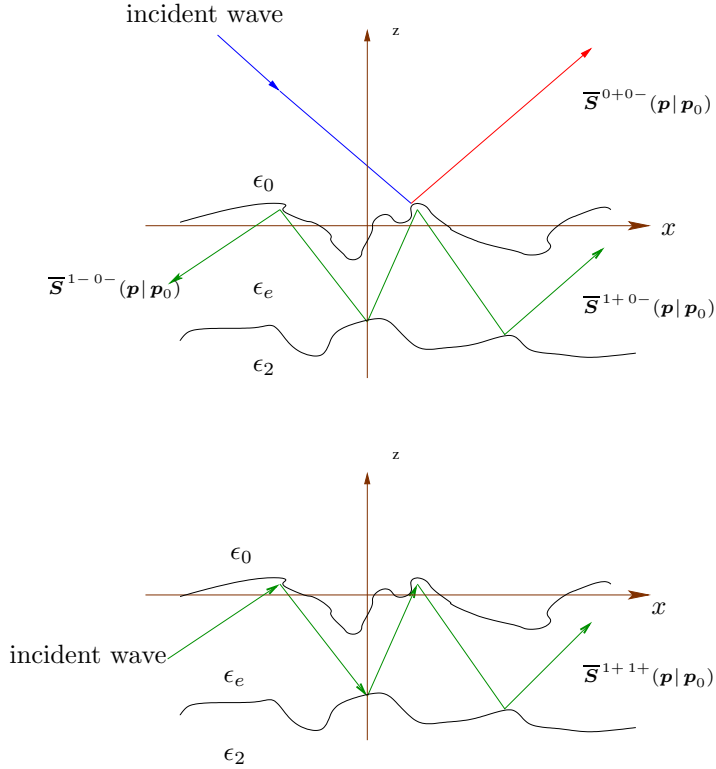,height=10cm}
      \caption{Scattering operator definitions.}
      \label{Chap3Fig3}
 \end{figure}

The basis $[\evp{0+}{},\ehp{}]$ and $[\evp{1-}{},\ehp{}]$ respectively orthogonal to the vectors
$\vec{k}_{\p{}}^{0+}$ and $\vec{k}_{\p{}}^{1-}$ are then defined by
\begin{eqnarray} \ehp{}& = \frac{\ez
\times \vec k_{\p{}}^{0\pm}}{||\ez \times \vec k_{\p{}}^{0\pm}||}=\ez \times \hvec{p} \, ,
\label{TE0}\\
\evp{0\pm}{} & =   \frac{\ehp{} \times \vec
k_{\p{}}^{0\pm}}{||\ehp{0} \times \vec k_{\p{}}^{0\pm}||}=\pm
\frac{\alpha_0(\vec p)}{K_0}\hvec{p}-\frac{||\vec p||} {K_0}\ez \,
, \label{TM0}
\end{eqnarray}
and
\begin{eqnarray}
\ehp{}& = \frac{\ez \times \vec k_{\p{}}^{1\pm}}{||\ez
\times \vec k_{\p{}}^{1\pm}||}=\ez \times \hvec{p} \, ,\label{TE1}\\
\evp{1\pm}{} & =   \frac{\ehp{} \times \vec
k_{\p{}}^{1\pm}}{||\ehp{0} \times \vec k_{\p{}}^{1\pm}||}=\pm
\frac{\alpha_e(\vec p)}{K_e}\hvec{p}- \frac{||\vec p||}{K_e}\ez \,
. \label{TM1}
\end{eqnarray}
The scattering operators can be written with dyadic notations on these bases:
\begin{eqnarray}
\fl\op{R}^{10}(\p{}|\p{0})=
R^{10}(\p{}|\p{0})_{VV}\,\,\hvec{e}^{0+}_{V}(\p{})\hvec{e}^{0-}_{V}(\p{0})+R^{10}(\p{}|\p{0})_{HV}\,\,\hvec{e}_{H}(\p{})\hvec{e}^{0-}_{V}(\p{0})\no\\
+
R^{10}(\p{}|\p{0})_{VH}\,\,\hvec{e}^{0+}_{V}(\p{})\hvec{e}_{H}(\p{0})+R^{10}(\p{}|\p{0})_{HH}\,\,\hvec{e}_{H}(\p{})\hvec{e}_{H}(\p{0})\,,\\
\fl\op{T}^{10}(\p{}|\p{0})=
T^{10}(\p{}|\p{0})_{VV}\,\,\hvec{e}^{1-}_{V}(\p{})\hvec{e}^{0-}_{V}(\p{0})+T^{10}(\p{}|\p{0})_{HV}\,\,\hvec{e}_{H}(\p{})\hvec{e}^{0-}_{V}(\p{0})\no\\
+
T^{10}(\p{}|\p{0})_{VH}\,\,\hvec{e}^{1-}_{V}(\p{})\hvec{e}_{H}(\p{0})+T^{10}(\p{}|\p{0})_{HH}\,\,\hvec{e}_{H}(\p{})\hvec{e}_{H}(\p{0})\,,
\end{eqnarray}
or in a matrix form:
\begin{eqnarray}
\left[\op{R}^{10}(\p{}|\p{0})\right] &=\left(\begin{array}{cc} R^{10}(\p{}|\p{0}) _{VV} & R^{10}(\p{}|\p{0}) _{VH} \\
R^{10}(\p{}|\p{0}) _{HV} & R^{10}(\p{}|\p{0}) _{HH} \end{array} \right)\,,\\
\left[\op{T}^{10}(\p{}|\p{0})\right] &=\left(\begin{array}{cc} T^{10}(\p{}|\p{0}) _{VV} & T^{10}(\p{}|\p{0}) _{VH} \\
T^{10}(\p{}|\p{0}) _{HV} & T^{10}(\p{}|\p{0}) _{HH}
\end{array}\right)\,.
\end{eqnarray}
In a similar way, we can define scattering operators $\op{R}^{01}$ and $\op{T}^{01}$ which describe the
reflected and transmitted fields when the source is in medium 1 (with permittivity $\ep_e$). For a rough surface
$z=h_2(\x)$ situated on the plane $z=-H$ separating two homogenous media with permittivity $\ep_e$ and $\ \ep_2$,
we introduce scattering operators $\op{R}^{H\,21}$ and $\op{T}^{H\,21}$  which describe the reflected field in
medium 1 and the transmitted  field in medium 2 when the source is in  medium 1 (with the permittivity $\ep_e$).
These scattering operators can be obtained from the scattering operators $\op{R}^{21}$ and $\op{T}^{21}$  for a
rough surface situated on the plane $z=0$ using the following properties of the scattering
operators~\cite{Voro}:
\begin{eqnarray}
 \op{R}^{H\,21}(\p{}|\p{0})=\e^{\rmi\,(\alp{e}{}+\alp{e}{0})H}\,\op{R}^{21}(\p{}|\p{0})\,,\\
\op{T}^{H\,21}(\p{}|\p{0})=\e^{\rmi\,(-\alp{2}{}+\alp{e}{0})H}\,\op{T}^{21}(\p{}|\p{0})\,.
\end{eqnarray}
In the rest of this paper, we suppose that we know the scattering operator expressions for $\op{R}^{10}$,
$\op{T}^{10}$, $\op{R}^{01}$, $\op{T}^{01}$, $\op{R}^{21}$, $\op{T}^{21}$. Several approximate theories, like
the small perturbation~\cite{Rice}, the Kirchhoff~\cite{Beckmann}, the small-slope approximation~\cite{Voro1},
the full-wave method~\cite{Bahar4}, the integral-equation method~\cite{Fung,Alvarez}, and others
theories~\cite{Ogilvy,Voro,DeSanto}, can be used to obtain expressions for these operators~\cite{Voro}. With
them, we can formally write the scattering operators for a slab with rough boundaries separating two homogeneous
media (see \Fref{Chap3Fig3}).
We use the following notation $\RpS{ab}{}{0}$ for these scattering operators. The upperscripts $a$ and $b$
indicate the receiver location, the source location and if the waves are upgoing or downgoing. For example, the
operator $\RpS{\,1+\,0-}{}{0}$ describes the amplitude of an incident downgoing wave from  medium 0 which is
scattered into an upgoing wave in medium 1. The upwell electric field in the medium 1 is given by
\begin{equation}
\E^{1+}(\x,z)=\intpV{}{} \,\e^{\rmi\,\p{}\cdot\x+\rmi\,\alp{e}{}z}\,\RpS{1+,0-}{}{0}\cdot
\E^{0\,i}(\p{0})\,,\label{ChampES1+0-}
\end{equation}
and
\begin{eqnarray}
\fl\RpS{1+,0-}{}{0}=&
S^{1+0-}(\p{}|\p{0})_{VV}\,\,\hvec{e}^{1+}_{V}(\p{})\hvec{e}^{0-}_{V}(\p{0})+S^{1+0-}(\p{}|\p{0})_{HV}\,\,
\hvec{e}_{H}(\p{})\hvec{e}^{0-}_{V}(\p{0})\no\\+&
S^{1+0-}(\p{}|\p{0})_{VH}\,\,\hvec{e}^{1+}_{V}(\p{})\hvec{e}_{H}(\p{0})
+S^{1+0-}(\p{}|\p{0})_{HH}\,\,\hvec{e}_{H}(\p{})\hvec{e}_{H}(\p{0})\,.
\end{eqnarray}
%On ajoute un signe $\pm$ aux ondes se propageant \`a l'int\'erieur de
%la couche car les ondes peuvent \^etre ascendantes o\`u
%descendantes.
%De m\^eme, on d\'efinira avec les m\^emes conventions les matrices de
%diffusion $\RpSe{\times \times}{}{0}$ o\`u le milieu $\ep_1$ est
%rempla\c{c}\'e par le milieu effectif $\ep_e$.
To express the different operators $\op{S}^{ab}$ in function of $\op{R}^{10}$, $\op{T}^{10}$, $\op{R}^{01}$,
$\op{T}^{01}$, $\op{R}^{21}$, $\op{T}^{21}$, we have just to add formally  all the multiple scattering
contributions on the rough boundaries~(\cite{Voro}, p.24). For example, $\RpS{1+\,0-}{}{0}$ is given by
\begin{eqnarray}
\fl \op{S}^{1+0-}=\op{R}^{H\,21}\cdot\op{T}^{10}+\op{R}^{H\,21}\cdot\op{R}^{01}\cdot\op{R}^{H\,21}\cdot\op{T}^{10}\no\\
+\op{R}^{H\,21}\cdot\op{R}^{01}\cdot\op{R}^{H\,21}\cdot\op{R}^{01}\cdot\op{R}^{H\,21}\cdot\op{T}^{10}+\dots\,\\
=\op{R}^{H\,21}\cdot[\op{I}^{1-1-}_{\perp}-\op{R}^{01}\cdot\op{R}^{H\,21}]^{-1}\cdot\op{T}^{10}\,,
\end{eqnarray}
where
%$\E^{0\,i-}$ is the amplitude of the incident field and
we have used the following notations:
\begin{equation}
[\op{A}\cdot\op{B}](\p{}|\p{0})=\intpV{1}{}\,\op{A}(\p{}|\p{1})\cdot\op{B}(\p{1}|\p{0})\,.\label{produitpp0}
\end{equation}
%and
%\begin{eqnarray}
%K^{r}_{e}&=\sqrt{\Re[\ep_e'(\omega)]}\,K_{vac}\,, \\
%\end{eqnarray}
We have defined the projectors $\op{I}_{\perp}^{1a1a_0}$ by
\begin{eqnarray}
\op{I}^{1a1a_0}_\perp(\p{})&= \evp{1\,a}{}\evp{1\,a_0}{}+\ehp{}\ehp{}\,,\\
\op{I}^{1a1a_0}_{\perp}(\p{}|\p{0})&=(2\pi)^2\,\delta(\p{}-\p{0})\,\op{I}^{1a1a_0}_\perp(\p{})\,,
\end{eqnarray}
where $a$ and  $a_0$ are the sign $+$ or $-$. The operator $\op{I}^{1a1a_0}_{\perp}(\p{}|\p{0})$ is the linear
identity mapping
 from the space vector   defined by the  basis $[\evp{1a_0}{}$,$\ehp{}]$ to the space vector defined by
$[\evp{1a}{}$,$\ehp{}]$. To write the electric field $\E^{1+}(\x,z)$ inside the slab with the scattering
operator $\op{S}^{1+0-}$, we have implicitly assumed that the layer is sufficiently thick in order to have the
condition $-H+max_{\x}h_2(\x)<z<min_{\x}h_1(\vec{x})$. Furthermore, we will suppose that all the particles are
inside the layer defined by $-H+max_{\x}h_2(\x)<z<min_{\x}h_1(\vec{x})$. Otherwise, the Rayleigh
hypothesis~\cite{Voro} must be invoked to justify the use of the scattering operator for $-H+max_{\x}h_2(\x)>z$
and $z>min_{\x}h_1(\vec{x})$. 
%At last, if the particles are not too closed to the boundaries (at least one
%wavelength), we can also neglect the interaction of the scatterers with the evanescent waves %propagating along
%the boundaries. In this case, we restrict the integration domain to $||\p{}||<K^r_e$ in  equation
%\eref{ChampES1+0-}. The limit $K^r_e$  separating  the space waves and the evanescent waves %is given by
%$K^r_e=\sqrt{\ep'_e}K_{vac}$ for $\ep_e=\ep_e'+\rmi\ep_e''$ if we suppose that $|\ep_e''|\ll|\ep_e'|$.

Using the same reasoning for the different contributions, we obtain:
\begin{eqnarray}
\RS{1-\,1-}&=[\op{I}^{1-1-}_{\perp}-\op{R}^{01}\cdot\op{R}^{H\,21}]^{-1}-\op{I}^{1-1-}_{\perp}\,,\label{S1-1-}\\
\RS{1+\,1-}&=\op{R}^{H\,21}\cdot[\op{I}^{1-1-}_{\perp}+\RS{1-\,1-}]\,,\label{S1+1-}\\
\RS{1-\,1+}&=[\op{I}^{1-1-}_\perp+\RS{1-\,1-}]\cdot\op{R}^{01}\,,\\
\RS{1+\,1+}&=[\op{I}^{1+1+}_\perp-\op{R}^{H\,21}\cdot\op{R}^{01}]^{-1}-\op{I}^{1+1+}_\perp\,,\\
%&=\op{R}^{H\,21}\cdot\op{R}^{01}+\op{R}^{H\,21}\cdot\RS{1-\,1-}\cdot\op{R}^{01}\,,\\
&=\op{R}^{H\,21}\cdot\RS{1-\,1+}\,,\\
\RS{0+\,0-}&=\op{R}^{10}+\op{T}^{01}\cdot\RS{1+\,1-}\cdot\op{T}^{10}\,,\label{S0+0-}\\
\RS{1+\,0-}&=\RS{1+\,1-}\cdot\op{T}^{10}\,,\\
\RS{1-\,0-}&=[\op{I}^{1-1-}_\perp+\RS{1-\,1-}]\cdot\op{T}^{10}\,,\\
\RS{0+\,1-}&=\op{T}^{01}\cdot\RS{1+\,1-}\,,\\
\RS{0+\,1+}&=\op{T}^{01}\cdot[\op{I}_\perp^{1+1+}+\RS{1+\,1+}]\,.
\end{eqnarray}
%Remarquons que suivant  la m\'ethode des \'equations de Rayleigh
%r\'eduites, nous pouvons obtenir directement des approximations des
%matrices $\RS{\times\,\times}$ dans le cas de la m\'ethode des %faibles
%perturbations, comme cela a \'et\'e pr\'esent\'e dans
%le chapitre \eref{Chapitre2}.
We now show how to express the Green operators
$\op{G}^{aa_0}_S(\rr,\rr_0)$ with the scattering operators
$\RpS{a\pm\,a_0\pm}{}{0}$.

To determine, for example, the Green operator $\op{G}_{S}^{1\,1}(\rr,\rr_0)$, we have to calculate the field
produced inside the slab by a spherical source which is also in medium 1. We can decompose the function
$\op{G}_{S}^{1\,1}(\rr,\rr_0)$ under the following form:
\begin{eqnarray}
 \op{G}_{S}^{1\,1}(\rr,\rr_0)=\op{G}_{1}^{\,\infty}(\rr,\rr_0)+\op{G}_{S}^{1+\,1+}(\rr,\rr_0)\no\\
+\op{G}_{S}^{1+\,1-}(\rr,\rr_0)+\op{G}_{S}^{1-\,1-}(\rr,\rr_0)+\op{G}_{S}^{1-\,1+}(\rr,\rr_0)\,.\label{contribGS11}
\end{eqnarray}
The first term is the spherical source term in an infinite homogeneous medium (with permittivity $\ep_e$), and
the following terms are the fields produced by multiple scattering process on the rough boundaries. We can
obtain these contributions noticing that the source term
$$\op{G}_{1}^{\,\infty}(\rr,\rr_0)=\left(\op{I}+\frac{1}{K_e^2}\nabla\nabla\right)P.V.\frac{\e^{\rmi\,K_e\,||\rr-\rr_0||}}{4\,\pi\,||\rr-\rr_0||}\,$$
can be decomposed as a linear combination of plane wave using the Weyl formula (\ref{AppA}):
\begin{eqnarray}
  \fl \mbox{for} \quad z>z_0 :\no \\
\fl\op{G}_{1}^{\,\infty}(\rr,\rr_0)=\frac{\rmi}{2}\,\intpV{0}{}\e^{\rmi\,\p{0}\cdot(\x-\x_0)
+\rmi\,\alp{e}{0}\,(z-z_0)}\,(\op{I}-\hvec{k}^{1\,+}_{\p{0}}\hvec{k}^{1\,+}_{\p{0}})\,\frac{1}{\alp{e}{0}}\,;\\
 \fl\mbox{for} \quad z<z_0 :  \no\\
\fl\op{G}_{1}^{\,\infty}(\rr,\rr_0)=\frac{\rmi}{2}\,\intpV{0}{}\e^{\rmi\,\p{0}\cdot(\x-\x_0)-
\rmi\,\alp{e}{0}\,(z-z_0)}\,(\op{I}-\hvec{k}^{1\,-}_{\p{0}}\hvec{k}^{1\,-}_{\p{0}})\,
\frac{1}{\alp{e}{0}}\,,\label{Chap3weylinf}
\end{eqnarray}
with $\vec{k}^{1\,\pm}_{\p{0}}= \p{0}\pm\alp{e}{0}\,\ez$ and $\hvec{k}^{1\,\pm}_{\p{0}}=\vec{k}^{1\,\pm}_{\p{0}}/||\vec{k}^{1\,\pm}_{\p{0}}||$. Hence, for $z \neq z_0$, $\op{G}_{1}^{\,\infty}$
is a linear combination of the following plane waves:
\begin{equation}
\frac{\rmi}{2\,\alp{e}{0}}\,\e^{\rmi\,\hvec{k}^{1\,\pm}_{\p{0}}\cdot(\rr-\rr_0)}(\op{I}-\hvec{k}^{1\,\pm}_{\p{0}}\hvec{k}^{1\,\pm}_{\p{0}})\,,
\end{equation}
which are  transverse to the propagation directions $\hvec{k}^{1\,\pm}_{\p{0}}$ because
$(\op{I}-\hvec{k}^{1\,\pm}_{\p{0}}\hvec{k}^{1\,\pm}_{\p{0}})\cdot \hvec{k}^{1\,\pm}_{\p{0}}=0$. For each of
these plane waves, we can calculate the field produced by the scattering process on the boundaries with the
scattering operators $\RS{1\times\,1\times}$. By using the superposition principle of the electric
field~\cite{Jackson}, we have for  the contribution due to $\op{S}^{1+\,1-}$:
\begin{eqnarray}
\fl \op{G}_{S}^{1+\,1-}(\rr,\rr_0)=\frac{\rmi}{2}\iintpV{}{0}{}
& \e^{\rmi\,\p{}\cdot\x-\rmi\,\p{0}\cdot\x_0+\rmi\,\alp{e}{}\,z+\rmi\,\alp{e}{0}\,z_0}\no\\
& \times \RpS{1+\,1-}{}{0}\cdot(\op{I}-\hvec{k}^{1\,-}_{\p{0}}\hvec{k}^{1\,-}_{\p{0}})\,\frac{1}{\alp{e}{0}}\,,\no\\
\end{eqnarray}
since in this case, the incident waves are propagating along the direction $\hvec{k}^{1\,-}_{\p{0}}$. We
summarize all the contributions \eref{contribGS11}
 to $\op{G}_{S}^{1\,1}$ under the form:
\begin{eqnarray}
\op{G}_{S}^{1\,1}(\rr,\rr_0)=\op{G}_{1}^{\,\infty}(\rr,\rr_0)+\sum_{a,a_0=\pm}\op{G}_{S}^{1a\,1a_0}(\rr,\rr_0)\,,\no\\\label{contribGS11-2}
\end{eqnarray}
with
\begin{eqnarray}
\fl\op{G}_{S}^{1a\,1a_0}(\rr,\rr_0)=\frac{\rmi}{2}\iintpV{}{0}{} &
\e^{\rmi\,\p{}\cdot\x-\rmi\,\p{0}\cdot\x_0+a\,\rmi\,\alp{e}{}\,z-a_0\,\rmi\,\alp{e}{0}\,z_0}\no\\
& \times
\RpS{1a\,1a_0}{}{0}\cdot(\op{I}-\hvec{k}^{1\,a_0}_{\p{0}}\hvec{k}^{1\,a_0}_{\p{0}})
\frac{1}{\alp{e}{0}}\,.\no\\
\end{eqnarray}
where $a$, $a_0$ are the sign  $+$ or $-$. By using the same
arguments, we demonstrate that:
\begin{eqnarray}
\op{G}_{S}^{10}(\rr,\rr_0)=\op{G}_{S}^{1+\,0-}(\rr,\rr_0)+\op{G}_{S}^{1-\,0-}(\rr,\rr_0)\,,\\
\op{G}_{S}^{01}(\rr,\rr_0)=\op{G}_{S}^{0+\,1+}(\rr,\rr_0)+\op{G}_{S}^{0+\,1-}(\rr,\rr_0)\,,\\
\op{G}_{S}^{00}(\rr,\rr_0)=\op{G}_{0}^{\,\infty}(\rr,\rr_0)+\op{G}_{S}^{0+\,0-}(\rr,\rr_0)\,,\label{decompG0}
\end{eqnarray}
with
\begin{eqnarray}
\fl \op{G}_{S}^{1a\,0-}(\rr,\rr_0)=\frac{\rmi}{2}\iintpV{}{0}{}
&\e^{\rmi\,\p{}\cdot\x-\rmi\,\p{0}\cdot\x_0+a\,\rmi\,\alp{e}{}\,z+\,\rmi\,\alp{0}{0}\,z_0}\no
\\ &\times\RpS{1a\,0-}{}{0}\cdot(\op{I}-\hvec{k}^{0\,-}_{\p{0}}\hvec{k}^{0\,-}_{\p{0}})\frac{1}{\alp{0}{0}}\,,\no \\\\
\fl\op{G}_{S}^{0+\,1a_0}(\rr,\rr_0)=\frac{\rmi}{2}\iintpV{}{0}{}
&\e^{\rmi\,\p{}\cdot\x-\rmi\,\p{0}\cdot\x_0+\rmi\,\alp{0}{}\,z-a_0\,\rmi\,\alp{e}{0}\,z_0}\no
\\ &\times \RpS{0+\,1a_0}{}{0}\cdot(\op{I}-\hvec{k}^{1\,a_0}_{\p{0}}\hvec{k}^{1\,a_0}_{\p{0}})\frac{1}{\alp{e}{0}}\,,\no\\\\
\fl\op{G}_{S}^{0+\,0-}(\rr,\rr_0)=\frac{\rmi}{2}\iintpV{}{0}{}
&\e^{\rmi\,\p{}\cdot\x-\rmi\,\p{0}\cdot\x_0+\,\rmi\,\alp{0}{}\,z+\,\rmi\,\alp{0}{0}\,z_0}\no
\\ &\times \RpS{0+\,0-}{}{0}\cdot(\op{I}-\hvec{k}^{0\,-}_{\p{0}}\hvec{k}^{0\,-}_{\p{0}})\frac{1}{\alp{0}{0}}\,.\no\\
\end{eqnarray}
%where $K_0^r=\sqrt{\ep_0'(\omega)}K_vac$ and $\ep_0=\ep'_0+\rmi\ep''_0$
\section{Lipmann-Schwinger equations and scattered field}
If we iterate equation \eref{Lip5e}, we obtain the following series for the function $\op{G}_{SV}^{11}$:
\begin{eqnarray}
\op{G}_{SV}^{11}&=\op{G}_{S}^{11}+\op{G}_{S}^{11}\cdot\op{V}^{11}\cdot\op{G}_{SV}^{11}\,,\label{Lip2bis}\\
&=\op{G}_{S}^{11}+\op{G}_{S}^{11}\cdot\op{V}^{11}\cdot\op{G}_{S}^{11}
+\op{G}_{S}^{11}\cdot\op{V}^{11}\cdot\op{G}_{S}^{11}\cdot\op{V}^{11}\cdot\op{G}_{S}^{11}+\dots\,\no\\
\label{Lip2bis2}
\end{eqnarray}
If we introduce the transition operator $\op{T}_{SV}^{11}$ by
\begin{equation}
\op{G}_{SV}^{11}=\op{G}_{S}^{11}+\op{G}_{S}^{11}\cdot\op{T}_{SV}^{11}\cdot\op{G}_{S}^{11}\label{opT}\,,
\end{equation}
we obtain  Lipmann-Schwinger equations in comparing the definition in \eref{opT} with the development in
\eref{Lip2bis2}:
\begin{eqnarray}
\op{T}_{SV}^{11} &=\op{V}^{11}+\op{V}^{11}\cdot\op{G}_{S}^{11}\cdot\op{T}_{SV}^{11}\,,\label{Chap3TeV}\\
\op{T}_{SV}^{11}
&=\op{V}^{11}+\op{T}^{11}_{SV}\cdot\op{G}_{S}^{11}\cdot\op{V}^{11}\,.
\end{eqnarray}
With these operators, we can straightforwdly rewrite in a compact form equations (\ref{Lip1e}-\ref{Lip6e}):
\begin{eqnarray}
\op{G}_{SV}^{00}&=\op{G}_{S}^{00}+\op{G}_{S}^{01}\cdot\op{T}_{SV}^{11}\cdot\op{G}_{S}^{10}\,,\label{Lip1T}\\
\op{G}_{SV}^{10}&=\op{G}_{S}^{10}+\op{G}_{S}^{11}\cdot\op{T}_{SV}^{11}\cdot\op{G}_{S}^{10}\,,\label{Lip2T}\\
\op{G}_{SV}^{20}&=\op{G}_{S}^{20}+\op{G}_{S}^{21}\cdot\op{T}_{SV}^{11}\cdot\op{G}_{S}^{10}\,,\label{Lip3T}\\
\op{G}_{SV}^{01}&=\op{G}_{S}^{01}+\op{G}_{S}^{01}\cdot\op{T}_{SV}^{11}\cdot\op{G}_{S}^{11}\,,\label{Lip4T}\\
\op{G}_{SV}^{21}&=\op{G}_{S}^{21}+\op{G}_{S}^{21}\cdot\op{T}_{SV}^{11}\cdot\op{G}_{S}^{11}\,.\label{Lip5T}
\end{eqnarray}
%As for equation (\ref{Lip1e},\ref{Lip6e}), we can verify that these equations are %solution of our problem.
All the scattering processes in the random medium are contained in the transition operator $\op{T}^{11}_{SV}$.
If we know this operator, then we can calculate all the fields in the different media  using
(\ref{Lip1T}-\ref{Lip5T}). Furthermore, for any source  $\vec{j}_{source}$ in the medium 0, the incident
electric field is
\begin{eqnarray}
\E^{0i}(\rr)&=\rmi\,\omega\,\mu_{vac}\int \rmd^3\rr_0
\,\op{G}_{0}^{\infty}(\rr,\rr_0)\cdot\vec{j}_{source}(\rr_0)\,,\label{Chap3ES0r}
\end{eqnarray}
where $\mu_{vac}$ is the vacuum permeability. The resulting field  in medium 0 is given from \eref{Lip1T} and
\eref{decompG0} by
\begin{eqnarray}
\E_{SV}^{0}&=\E^{0i}+\E^{0s}_{SV}\,,\\
\E^{0s}_{SV}&=\E^{0s}_{S}+\op{G}_{S}^{01}\cdot\op{T}^{11}_{SV}\cdot\E_{S}^{1\,t}\,,\label{Chap3Eref}
\end{eqnarray}
where
\begin{eqnarray}
\E_{S}^{0s}(\rr)&=\rmi\,\omega\,\mu_{vac}\int \rmd^3\rr_0
\,\op{G}_{S}^{0+0-}(\rr,\rr_0)\cdot\vec{j}_{source}(\rr_0)\,,\label{Chap3ES0r-2}\\
\E_{S}^{1\,t}(\rr)&=\rmi\,\omega\,\mu_{vac}\int \rmd^3\rr_0
\,\op{G}_{S}^{10}(\rr,\rr_0)\cdot\vec{j}_{source}(\rr_0)\,.\label{Chap3ESet}
\end{eqnarray}
The field $\E_{SV}^{0s}$ is the scattered field produced by the random medium and the rough
boundaries which is decomposed in two contributions where $\E_{S}^{0s}(\rr)$ is the reflected field produced by an homogeneous slab (with permittivity $\ep_e$) with
rough boundaries, and the second  term $\op{G}_{S}^{01}\cdot\op{T}^{11}_{SV}\cdot\E_{S}^{1\,t}$ is the field
scattered by  the random medium and  the rough surfaces. Vector $\E_{S}^{1\,t}$ is the transmitted field in
medium 1 before any scattering by  the particles.
\section{Ensemble average} In order to calculate the coherent
field reflected by the random slab, we need to define the averaging procedure. Let $<\cdots>_S$ and $<\cdots>_V$
denote, respectively, the ensemble average over the surfaces  and the volume disorder. We also denote
$\ll\cdots\gg_{SV}$ the average over the rough surfaces and the volume disorder. We suppose that the rough
surfaces and random medium  properties are statistically \emph{independent}, which means that
$\ll\cdots\gg_{SV}=<[<\cdots>_S]>_V=<[<\cdots>_V]>_S$. In the following development, we won't use any specific
statistical properties of the rough surfaces (except that $<h_1(\x)>_S=<h_2(\x)>_S=0$); thus, we don't need to specify them. An extensive description can
be found in the references \cite{Ogilvy,Bennett}. The ensemble average over the random medium is defined by
\begin{equation}
<f>_V=\int_{V_1}\rmd^3\rr_1\dots\rmd^3\rr_N\,\,f(\rr_1,\dots,\rr_N)\,p(\rr_1,\dots,\rr_N)\,,
\end{equation}
where $\rr_1,\cdots,\rr_N$ are the particles positions, and $p(\rr_1,\dots,\rr_N)$ is the probability density
function of finding the N particles at positions $\rr_1,\cdots,\rr_N$. We will use a decomposition of this
density function with conditional probabilities~\cite{Kong,Kong2001-3}:
\begin{eqnarray}
p(\rr_1,\dots,\rr_N)=p(\rr_i)p(\rr_1,\cdots,\widehat{\rr_{i}},\cdots\rr_{N}|\rr_i)\,,\label{defp1}\\
p(\rr_1,\dots,\rr_N)=p(\rr_i)p(\rr_j|\rr_i)p(\rr_1,\cdots,\widehat{\rr_{i}},\cdots\widehat{\rr_j},\cdots,\rr_{N}|\rr_i,\rr_j)\,,\label{defp2}
\end{eqnarray}
where the hat $\,\widehat{}\,$ indicates that the term is absent. The function $p(\rr_i)$ is the probability
density function of finding a particle at $\rr_i$, $p(\rr_i|\rr_j)$ is the conditional probability of finding a
particle at $\rr_i$ given a particle at $\rr_j$, etc. If the particles are uniformly distributed inside the
random medium $V_1$, then the single particle density function is $p(\rr_i)=1/\mathcal{V}_1$, where
$\mathcal{V}_1$ is the volume of the area $V_1$. In this case, we  also define a pair-distribution function by
\begin{equation}
g(||\rr_i-\rr_j||)=p(\rr_i|\rr_j)/p(\rr_i)\,,
\end{equation}
which depends only on the distance between the two particles if we suppose that the  distribution of the
particles is statistically homogeneous and isotropic. The normalization factor $\mathcal{V}_1$ is chosen in such
a way that when the particles located at $\rr_i$, $\rr_j$ are far away from each over, their positions are
uncorrelated (i.e., $p(\rr_i|\rr_j)=p(\rr_i)$), and we have
\begin{equation}
\lim_{||\rr_i-\rr_j||\to+\infty}g(||\rr_i-\rr_j||)=1\,.
\end{equation}
With these conditional probability functions, we define
conditional averages:
\begin{eqnarray}
\fl<f>_{V;\rr_i}=\int_{V_1}\rmd^3\rr_1\dots\widehat{\rmd^3\rr_i}\dots\rmd^3\rr_N\,\,f(\rr_1,\dots,\rr_i,\dots,\rr_N)\,\no
\\
\times p(\rr_1,\dots,\widehat{\rr_i},\dots\rr_N|\rr_i)\,,\label{avep1}\\
\fl<f>_{V;\rr_i,\rr_j}=\int_{V_1}\rmd^3\rr_1\dots\widehat{\rmd^3\rr_i}\dots\widehat{\rmd^3\rr_j}\dots\rmd^3\rr_N\,\,f(\rr_1,\dots,\rr_i,\dots,\rr_j,\dots,\rr_N)\,\\
\times
\,p(\rr_1,\dots,\widehat{\rr_i},\dots,\widehat{\rr_j},\dots,\rr_N|\rr_i,\rr_j)\,.\label{avep2}
\end{eqnarray}
\section{Coherent potential approximation and effective medium theory}
\label{secCPA}
Until now, we have not clarified how to determine the effective permittivity $\ep_e(\omega)$. To determine
$\ep_e(\omega)$, we use the fact that  under some assumptions (mainly that the effective medium is not spatially
dispersive~\cite{Sheng1} i.e., $\ep(\omega,\vec{k})=\ep(\omega)$), it can be shown, using a diagrammatic
technique, that the coherent part of the field $<\op{E}>_{V}$ which propagates inside an infinite  random medium
behaves as a wave in an homogeneous medium  with a renormalized effective
permittivity~\cite{Kong,Kong2001-3,Sheng1,Sheng2,Apresyan,Ishi1,Frish}. In order to insure this result in a
self-consistent way, we introduce the Coherent Potential Approximation (CPA), which postulates
that~\cite{Kong,Kong2001-3,Sheng1,Soven,Korringa}
\begin{equation}
<\op{G}_{SV}^{11}>_V=\op{G}_{S}^{11}\,.\label{CPAcond}
\end{equation}
This equation is in fact the definition of our effective permittivity $\ep_e$ and is a generalization of the
classical (CPA) approach since we take into account the boundaries in the Green function definitions. Using
equation \eref{opT}, we immediately see that condition \eref{CPAcond} is equivalent to
\begin{equation}
<\op{T}_{SV}^{11}>_V=\op{0}\,,\label{PropCPA}
\end{equation}
where
\begin{equation}
\op{T}_{SV}^{11} =\op{V}^{11}+\op{V}^{11}\cdot\op{G}_{S}^{11}\cdot\op{T}_{SV}^{11}\,.\label{TSV11_2}
\end{equation}
Therefore, equations (\ref{PropCPA}, \ref{TSV11_2}) provide a closed system of equations on the unknown
$\ep_e(\omega)$ which takes place in the definition of $\op{V}^{11}$ and $\op{G}_S^{11}$. Because the random
medium is made with spherical particles, it is convenient to express equation \eref{TSV11_2} as a function of
the scattering operator $\op{t}^{11}_{\rr_i}$ for one particle located at $\rr_i$ in a infinite medium
(\ref{AppB}). It is defined by
\begin{equation}
\op{t}^{11}_{\rr_i}=[\op{I}-\op{v}^{11}_{\rr_i}\cdot\op{G}_{1}^{\infty}]^{-1}\cdot\op{v}^{11}_{\rr_i}\,,\label{deft11}
\end{equation}
where the scattering potential $\op{v}^{11}_{\rr_i}$ is:
\begin{eqnarray}
\op{v}^{11}_{\rr_i}(\rr,\rr_0)&=&(2\pi)^2\,\delta(\rr-\rr_0)\op{v}^1_{\rr_i}(\rr)\,,\label{Defv11}\\
\op{v}^1_{\rr_i}(\rr)&=&K_{vac}^2(\ep_s-\ep_1)\Theta_s(\rr-\rr_i)\op{I}\,.\label{Defv1}%&=& K_{vac}^2(\tilde\ep_s-\ep_e)
\end{eqnarray}
%with $\tilde\ep_s=\ep_s-(\ep_1-\ep_e)$.
%Since we have introduce an effective permittivity, the
In order to write  equation \eref{TSV11_2} with the operators $\op{t}^{11}_{\rr_i}$, following the Korringa
demonstration~\cite{Korringa}, we decompose
\begin{equation}
\op{V}^{11}(\rr,\rr_0)=(2\pi)^2\,\delta(\rr-\rr_0)\,K^2_{vac}(\ep_V(\rr)-\ep_e)\,\op{I}
\label{expV11epsV}\end{equation}
 and
\begin{equation}
\op{T}^{11}_{SV}=\op{V}^{11}+\op{V}^{11}\cdot\op{G}_{S}^{11}\cdot\op{T}^{11}_{SV} \end{equation}
 under the
following form:
\begin{equation}
\optilde{V}^{11}=\op{V}^{11}+\op{W}^{11}\,,\label{vtilde}
\end{equation}
where
\begin{equation}
\op{W}^{11}(\rr,\rr_0)=(2\pi)^2\,\delta(\rr-\rr_0)\,K^2_{vac}(\ep_e-\ep_1)\,\op{I}\,,\label{defW11}
\end{equation}
and
\begin{equation}
\optilde{T}^{11}_{SV}=\op{T}^{11}_{SV}+\op{Q}^{11}_{SV}\,,\label{defQ11}
\end{equation}
with
\begin{equation}
\op{Q}^{11}_{SV}=\op{W}^{11}+\op{W}^{11}\cdot\op{G}_{S}^{11}\cdot\op{T}^{11}_{SV}\,.\label{expQ11}
\end{equation}
By using the definitions (\ref{vtilde}, \ref{defW11}, \ref{expV11epsV}, \ref{Defv11}, \ref{Defv1}) we obtain
\begin{equation}
\optilde{V}^{11}=\sum_{i=1}^N\op{v}^{11}_{\rr_i}\,,\label{sumv11}
\end{equation}
and in inserting \eref{vtilde} in \eref{TSV11_2} and using the definition (\ref{defQ11}, \ref{expQ11}), we have
\begin{eqnarray}
\optilde{T}^{11}_{SV}&=\optilde{V}^{11}+\optilde{V}^{11}\cdot\op{G}_{S}^{11}\cdot\op{T}^{11}_{SV}
\\ &=\optilde{V}^{11}+\optilde{V}^{11}\cdot\op{G}_{S}^{11}\cdot(\optilde{T}^{11}_{SV}-\op{Q}^{11}_{SV})\,.\label{expT11-2}
\end{eqnarray}
In combining equation \eref{sumv11} and \eref{expT11-2}, we decompose $\optilde{T}^{11}_{SV}$ in the following
form:
\begin{equation}
\optilde{T}^{11}_{SV}=\sum_{i=1}^N\,\op{C}^{11}_{SV,\rr_i}\,,
\end{equation}
where
\begin{equation}
\op{C}^{11}_{SV,\rr_i}=\op{v}^{11}_{\rr_i}+\op{v}_{\rr_i}^{11}\cdot\op{G}_{S}^{11}\cdot(\sum_{j=1}^N\,
\op{C}^{11}_{SV,\rr_j}-\op{Q}^{11}_{SV})\,.\label{C11-1}
\end{equation}
If we subtract $\op{v}_{\rr_i}^{11}\cdot\op{G}_{S}^{11}\cdot\op{C}^{11}_{SV;\rr_i}$ in both sides of equation
\eref{C11-1}, we obtain
\begin{equation}
(\op{I}-\op{v}_{\rr_i}^{11}\cdot\op{G}_{S}^{11})\cdot\op{C}^{11}_{SV;\rr_i}=\op{v}^{11}_{\rr_i}+\op{v}_{\rr_i}^{11}\cdot\op{G}_{S}^{11}\cdot(\sum_{j=1,\,j\neq
i}^N\,\op{C}^{11}_{SV;\rr_j}-\op{Q}^{11}_{SV})\,,
\end{equation}
and then,
\begin{equation}
\op{C}^{11}_{SV,\rr_i}=\op{t}^{11}_{S,\rr_i}+\op{t}_{S,\rr_i}^{11}\cdot\op{G}_{S}^{11}\cdot(\sum_{j=1,\,j\neq
i}^N\,\op{C}^{11}_{SV,\rr_j}-\op{Q}^{11}_{SV})\,,\label{defC11SV}
\end{equation}
where
\begin{eqnarray}
\op{t}^{11}_{S,\rr_i}&=[\op{I}-\op{v}_{\rr_i}^{11}\cdot\op{G}_{S}^{11}]^{-1}\cdot\op{v}_{\rr_i}^{11}\,,\\
&=\op{v}^{11}_{\rr_i}+\op{v}^{11}_{\rr_i}\cdot\op{G}_{S}^{11}\cdot\op{t}^{11}_{S,\rr_i}\label{Deft11S}
\end{eqnarray}
is the scattering operator for only one  particle located at $\rr_i$ inside the slab. Using the decomposition
in \eref{contribGS11-2}:
\begin{equation}
\op{G}_{S}^{11}=\op{G}_{1}^{\infty}+\delta\op{G}_{S}^{11}\,,
\end{equation}
where
\begin{equation}
\delta\op{G}_{S}^{11}=\sum_{a,a_0=\pm}\,\op{G}_{S}^{1a1a_0}\,.
\end{equation}
In definition \eref{Deft11S}, we have
\begin{equation}
[\op{I}-\op{v}^{11}_{\rr_i}\cdot\op{G}_{1}^{\infty}]\cdot\op{t}^{11}_{S,\rr_i}=
\op{v}^{11}_{\rr_i}+\op{v}^{11}_{\rr_i}\cdot\delta\op{G}_{S}^{11}\cdot\op{t}^{11}_{S,\rr_i}\,,\label{eqDeft11S}
\end{equation}
and we demonstrate that
\begin{equation}
\op{t}^{11}_{S,\rr_i}=\op{t}^{11}_{\rr_i}
+\op{t}^{11}_{\rr_i}\cdot\delta\op{G}_{S}^{11}\cdot\op{t}^{11}_{S,\rr_i}\,,\label{deft11S-2}
\end{equation}
where
\begin{equation}
\op{t}^{11}_{\rr_i}=[\op{I}-\op{v}^{11}_{\rr_i}\cdot\op{G}_{1}^{\infty}]^{-1}\cdot\op{v}^{11}_{\rr_i}\,,
\end{equation}
which is equivalent to
\begin{equation}
\op{t}^{11}_{\rr_i}=\op{v}^{11}_{\rr_i}+\op{v}^{11}_{\rr_i}\cdot\op{G}_{1}^{\infty}\cdot\op{t}^{11}_{\rr_i}\,.
\end{equation}
\begin{figure}[htbp]
   \centering
   		\psfrag{z}{$z$}
      \psfrag{t}{$\op{t}^{11}_{\rr_i}(\rr,\rr_0)$}
      \psfrag{tS}{$\op{t}^{11}_{S,\rr_i}(\rr,\rr_0)$}
      \psfrag{ri}{$\rr_i$}
      \psfrag{C11}{$\op{C}^{11}_{SV,\rr_i}(\rr,\rr_0)$}
      \psfrag{H}{$H$}
      \psfrag{m0}{$\ep_0$}
      \psfrag{m1}{$\ep_e$}
      \psfrag{m2}{$\ep_2$}
      \psfrag{es}{$\ep_s$}
      \psfrag{r}{$\rr$}
      \psfrag{r0}{$\rr_0$}
         \epsfig{file=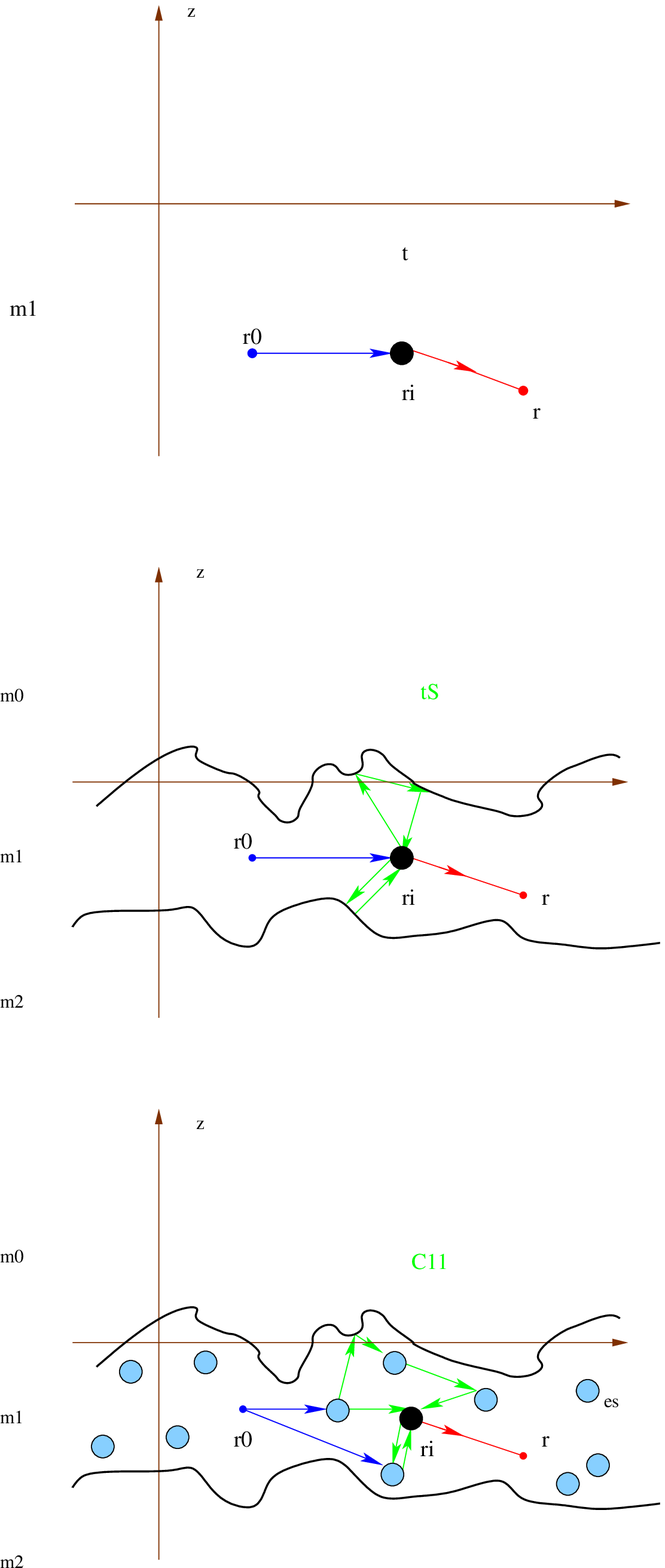,width=8cm}
         \caption{\label{Fig6}Graphical representation of $\op{t}^{11}_{\rr_i}$,
         $\op{t}^{11}_{S,\rr_i}$,$\op{C}^{11}_{SV,\rr_i}$}.
         \end{figure}
The operator $\op{t}^{11}_{\rr_i}$ describes the scattering process by a particle in an infinite homogenous
medium (Figure \ref{Fig6}). We must be careful in  determining the permittivity of this particle. In fact, as
the propagator between two scattering events inside the particle is $\op{G}_{1}^{\infty}$, the homogeneous
medium surrounding the particle has the permittivity $\ep_e$ due to the definition of $\op{G}_{1}^{\infty}$
(\ref{AppB}). However, $\op{t}^{11}_{\rr_i}$ doesn't describe the scattering by a particle of permittivity
$\ep_s$ inside a medium of permittivity $\ep_e$. If this was correct, the operator $\op{v}^{11}_{\rr_i}$ defined
by (\ref{Defv11}, \ref{Defv1}) would contain the factor $\ep_s-\ep_e$ and not the factor $\ep_s-\ep_1$. Thus, we
have to renormalize the particle permittivity in introducing $\tilde\ep_s=\ep_s-(\ep_1-\ep_e)$, such as
\begin{eqnarray}
\op{v}^1_{\rr_i}(\rr)&=K_{vac}^2(\ep_s-\ep_1)\Theta_s(\rr-\rr_i)\op{I}\,,\label{Defv1-2}\\
&=K_{vac}^2(\tilde\ep_s-\ep_e)\Theta_s(\rr-\rr_i)\op{I}\,.
\end{eqnarray}
Hence, the operator $\op{t}^{11}_{\rr_i}$ is the scattering operator for a single particle of permittivity
$\tilde\ep_s$ surrounded by an infinite medium of permittivity $\ep_e$.

The operator $\op{t}^{11}_{S,\rr_i}$  describes the scattering process by a particle located at $\rr_i$ inside
the volume $V_1$ of our slab (Figure \ref{Fig6}). If we iterate equation \eref{deft11S-2},
\begin{equation}
\fl
\op{t}^{11}_{S,\rr_i}=\op{t}^{11}_{\rr_i}+\op{t}^{11}_{\rr_i}\cdot\delta\op{G}_{S}^{11}\cdot\op{t}^{11}_{\rr_i}
+\op{t}^{11}_{\rr_i}\cdot\delta\op{G}_{S}^{11}\cdot\op{t}^{11}_{\rr_i}\cdot\delta\op{G}_{S}^{11}
\cdot\op{t}^{11}_{\rr_i}+\dots\,,\label{deft11S-3}
\end{equation}
we see that the first term is the scattering process due to the particle, and the following terms describe the
interaction between the particle and the boundaries  (Figure \ref{Fig6}) since the terms $\delta\op{G}_{S}^{11}$
come from the slab surfaces.

If we now look at equation \eref{defC11SV}, we see that it describes multiple scattering process by different
scatterers inside the slab. In fact, if we had defined the Green function $\op{G}^{11}_{S}$ without introducing
the effective medium $\ep_e(\omega)$ but in taking the permittivity $\ep_1$, the operators $\op{W}^{11}$ and
$\op{Q}^{11}_{SV}$ would have been  zero operators (since $\ep_e(\omega)=\ep_1(\omega)$ in the definition
(\ref{defW11}, \ref{expQ11}) give null contributions), and the iteration of equation \eref{defC11SV} show us the
multiple scattering process inside the slab~\cite{Sheng1,Frish}:
\begin{equation}
\fl\op{C}^{11}_{SV,\rr_i}=\op{t}^{11}_{S,\rr_i}+\sum_{j=1,\,j\neq i}^N\,\op{t}_{S,\rr_i}^{11}
\cdot\op{G}_{S}^{11}\cdot\op{t}^{11}_{S,\rr_j}\no\\
+\sum_{j=1,\,j\neq i}^N\,\sum_{k=1,\,k\neq j}^N\,\op{t}_{S,\rr_i}^{11}\cdot\op{G}_{S}^{11}
\cdot\op{t}^{11}_{S,\rr_j}\cdot\op{G}_{S}^{11}\cdot\op{t}^{11}_{S,\rr_k}+\dots\,.\no\\
\label{defC11SV-4}
\end{equation}
The operator $\op{C}^{11}_{SV,\rr_i}$ represents the field scattered by a particle located at $\rr_i$  which
takes into account all the interaction effects with the other particles and the boundaries. In introducing the
effective medium $\ep_e(\omega)$ in the definition of $\op{G}^{11}_{S}$, we see in equation \eref{defC11SV} that
the  multiple scattering contributions are attenuated by the factor $\op{Q}^{11}_{SV}$.

If we now average equation \eref{defQ11}, using the definitions (\ref{expQ11}, \ref{defW11}) and the (CPA)
hypothesis \eref{PropCPA}, we obtain
\begin{equation}
\fl (2\pi)^2\,\delta(\rr-\rr_0)\,\ep_e\,K_{vac}^2\,\op{I}=(2\pi)^2\,\delta(\rr-\rr_0)\,\ep_1\,K_{vac}^2\,\op{I}
+<\optilde{T}_{SV}^{11}(\rr,\rr_0)>_V\,,
\end{equation}
and since the ensemble average can be decomposed with the conditional probabilities \eref{defp1} and definition
\eref{avep1}, we have
\begin{eqnarray}
<\optilde{T}_{SV}^{11}>_V &=\sum_{i=1}^N\,<\op{C}^{11}_{SV,\rr_i}>_V\,,\\
&=\sum_{i=1}^N\,\int_{V_1}\rmd^3\rr_i\,p(\rr_i)\,<\op{C}^{11}_{SV,\rr_i}>_{V;\rr_i}\,,\\
&=n\,\int_{V_1}\rmd^3\rr_i\,<\op{C}^{11}_{SV,\rr_i}>_{V;\rr_i}\,,\label{lasteq}
\end{eqnarray}
where we have introduced the particles density $n=N/\mathcal{V}_1$. To obtain equation \eref{lasteq}, we have
used the fact that $<\op{C}^{11}_{SV;\rr_i}>_{V;\rr_i}=<\op{C}^{11}_{SV;\rr_j}>_{V;\rr_j}$ for $i\neq j$, since
we consider a statistical homogeneous random medium.

In averaging  equation \eref{defC11SV} with the conditional average $<\dots>_{V;\rr_i}$, and using  definitions
(\ref{defp2}, \ref{avep2}) we obtain:
\begin{eqnarray}
\fl <\op{C}^{11}_{SV,\rr_i}>_{V;\rr_i}=<\op{t}^{11}_{S,\rr_i}>_{V;\rr_i}
+\sum_{j=1,\,j\neq i}^N\int_{V_1}\rmd^3\rr_j p(\rr_j|\rr_i)\,\op{t}_{S,\rr_i}^{11}
\cdot\op{G}_{S}^{11}\cdot<\op{C}^{11}_{SV,\rr_j}>_{V;\rr_i,\rr_j}\,\no\\
\lo -<\op{t}_{S,\rr_i}^{11}\cdot\op{G}_{S}^{11}\cdot\op{Q}^{11}_{SV}>_{V;\rr_i}\,. \label{eqsurC11}
\end{eqnarray}
Since $\op{t}_{S;\rr_i}^{11}$ is the scattering operator for one particle located at $\rr_i$, it only depends on
the variable $\rr_i$ and not on $\rr_j$ with $j\neq i$, and the average $<\cdots>_{V;\rr_i}$ doesn't act on
$\op{t}_{S;\rr_i}^{11}$. Furthermore, the averaging of equation \eref{expQ11} is
\begin{equation}
<\op{Q}^{11}_{SV}>_{V;\rr_i}=\op{W}^{11}+\op{W}^{11}\cdot\op{G}_{S}^{11}\cdot<\op{T}^{11}_{SV}>_{V;\rr_i}\,.\label{aveQ11}
\end{equation}
This equation is simplified by  the (CPA) condition
$<\op{T}^{11}_{SV}>_V=\op{0}$, which can also be written under the
following form:
\begin{equation}
\int_{V_1}\rmd^3\rr_i\,p(\rr_i)\,<\op{T}^{11}_{SV}>_{V;\rr_i}=0\,.\label{identT11SV}
\end{equation}
As the identity in \eref{identT11SV} is valid whatever the volume $V_1$ is, we have
$<\op{T}^{11}_{SV}>_{V;\rr_i}=\vec{0}$. Thus, from equation (\ref{aveQ11}), 
we have:
\begin{equation}
<\op{Q}^{11}_{SV}>_{V;\rr_i}=\op{W}^{11}\,.\label{resul1}
\end{equation}
From 
the definition (\ref{defQ11},\ref{expQ11}) and the coherent potential approximation, we deduce that:
\begin{eqnarray}
<\op{Q}^{11}_{SV}>_V &=\op{W}^{11}+\op{W}^{11}\cdot\op{G}_{S}^{11}\cdot<\op{T}^{11}_{SV}>_V\,,\\
&=\op{W}^{11}\,.\label{resul2}
\end{eqnarray}
and 
\begin{eqnarray}
<\optilde{T}_{SV}^{11}>_{V}&=<\optilde{T}_{SV}^{11}>_{V}+<\op{Q}^{11}_{SV}>_V\,,\\
&=<\op{Q}^{11}_{SV}>_V\,.\label{resul3}
\end{eqnarray}
Accordingly, from the results (\ref{resul1}, \ref{resul2}, \ref{resul3}) and equation \eref{lasteq} we get for $i\in[1,N]$:
\begin{eqnarray}
<\op{Q}^{11}_{SV}>_{V;\rr_i}
&=<\optilde{T}_{SV}^{11}>_{V}\\
&=n\,\int_{V_1}\rmd^3\rr_j\,<\op{C}^{11}_{SV,\rr_j}>_{V;\rr_j}\,,
\end{eqnarray}
and equation \eref{eqsurC11} is 
\begin{eqnarray}
\fl <\op{C}^{11}_{SV,\rr_i}>_{V;\rr_i}=& \op{t}^{11}_{S,\rr_i}+n\,\op{t}_{S,\rr_i}^{11}\cdot
\op{G}_{S}^{11}\cdot\int_{V_1}\rmd^3\rr_j \left[g(||\rr_j-\rr_i||)\,<\op{C}^{11}_{SV,\rr_j}>_{V;\rr_i,\rr_j}\right.\no\\
&\left.\quad -<\op{C}^{11}_{SV,\rr_j}>_{V;\rr_j}\right]\,,
\label{eqsurC11_2}
\end{eqnarray}
where we have used 
%the properties that for $i\neq k$,
%\begin{equation}
%<\op{C}^{11}_{SV,\rr_j}>_{V;\rr_i,\rr_j}=<\op{C}^{11}_{SV,\rr_k}>_{V;\rr_i,\rr_k}
%\end{equation}
%for a statistical homogeneous medium, and 
the approximation $n \simeq (N-1)/\mathcal{V}_1$ which is valid for
 a large number of particle (${N}\gg 1$).
Using the same procedure, we can average equation \eref{defC11SV} with the conditional average
$<\dots>_{V;\rr_i,\rr_j}$ and obtain an equation on $<\op{C}^{11}_{SV,\rr_j}>_{V;\rr_i,\rr_j}$ in a function of
$<\op{C}^{11}_{SV,\rr_k}>_{V;\rr_i,\rr_j,\rr_k}$ and so on. Hence, a hierarchical system of equations can be
generated on the unknown $<\op{C}^{11}_{SV,\rr_i}>_{V;\rr_i}$, $<\op{C}^{11}_{SV,\rr_j}>_{V;\rr_i,\rr_j}$,
$<\op{C}^{11}_{SV,\rr_k}>_{V;\rr_i,\rr_j,\rr_k}$,\dots. We close this system by using the Quasi-Crystalline
Approximation (QCA) which states that~\cite{Kong,Kong2001-3,Gyorffy,Korringa,Lax}
\begin{equation}
<\op{C}^{11}_{SV,\rr_j}>_{V;\rr_i,\rr_j}=<\op{C}^{11}_{SV,\rr_j}>_{V;\rr_j}\,.
\end{equation}
As was demonstrated by Lax~\cite{Lax}, this approximation is strictly valid when the particles have a fixed position, as in a crystal. The quasi-crystalline approximation is equivalent to neglect of the fluctuation of the
effective field, acting on a particle located at $\rr_j$, due to a deviation of a particle located at $\rr_i$
from its average position.  Under this approximation, the effective permittivity $\ep_e(\omega)$ satisfies the
following system of equations:
\begin{eqnarray}
\fl(2\pi)^2\delta(\rr-\rr_0)\,\ep_e\,K_{vac}^2\,\op{I}=(2\pi)^2\delta(\rr-\rr_0)\,\ep_1\,K_{vac}^2\,\op{I}+n\,
\int_{V_1}\rmd^3\rr_i\,<\op{C}^{11}_{SV,\rr_i}(\rr,\rr_0)>_{V;\rr_i}\,,\no\\
\label{eff-1}
\\\fl <\op{C}^{11}_{SV,\rr_i}>_{V;\rr_i}=\op{t}^{11}_{S,\rr_i}+n\,\op{t}_{S,\rr_i}^{11}
\cdot\op{G}_{S}^{11}\cdot\int_{V_1}\rmd^3\rr_j
[g(||\rr_j-\rr_i||)-1]\,<\op{C}^{11}_{SV,\rr_j}>_{V;\rr_j}\,.\no\\\label{C11SVr}
\end{eqnarray}
We can simplify these equations by noticing that the contribution $\delta \op{G}^{11}_{S}$ due to the boundary
in $\op{G}^{11}_{S}=\op{G}^{\infty}_{1}+\delta \op{G}^{11}_{S}$ can be neglected in equations \eref{C11SVr} and
\eref{deft11S-2} if the following condition $K''_{e}H \gg 1$ with $K_e''=Im K_e$ is satisfied. Usually, we
define the extinction length as $l_e=1/2 K''_e$ and we see that the previous condition means that the slab
thickness must be greater than the extinction length.
%In fact, for a low concentration of scatterers ($n\ll 1$), the main contribution in the
%equation \eref{C11SVr} is given by the lowest order term in $n$ which is
%$\op{t}^{11}_{S,\rr_i}=\op{t}^{11}_{\rr_i}+\op{t}^{11}_{\rr_i}\cdot\delta
%\op{G}^{11}_{S}\cdot\op{t}^{11}_{S,\rr_i}$

For example, if we analyze the contribution $\op{G}_{S}^{1+\,0-}$ of $\delta\op{G}^{11}_{S}$, we have
\begin{eqnarray}
\fl\op{G}_{S}^{1+\,0-}(\rr,\rr_0)=\frac{\rmi}{2}\iintpV{}{0}{}
&\e^{\rmi\,\p{}\cdot\x-\rmi\,\p{0}\cdot\x_0+\rmi\,\alp{e}{}\,z+\,\rmi\,\alp{0}{0}\,z_0}\no
\\ &\times\RpS{1+\,0-}{}{0}\cdot(\op{I}-\hvec{k}^{0\,-}_{\p{0}}\hvec{k}^{0\,-}_{\p{0}})\frac{1}{\alp{0}{0}}\,,\no\\
\end{eqnarray}
where
\begin{eqnarray}
\fl\RpS{1+\,0-}{}{0}=& \iintpV{1}{2}{} \e^{+\rmi(\alp{e}{}+\alp{e}{1})\,H}\,\no\\
& \times\op{R}^{21}(\p{}|\p{1})\cdot\left[\op{I}_{\perp}^{1-1-}(\p{1}|\p{2})+\RpS{1-1-}{1}{2}\right]\cdot\op{T}^{10}(\p{2}|\p{0})\,,\no\\
\end{eqnarray}
where we have used the following property:
\begin{equation}
\op{R}^{H\,21}(\p{}|\p{1})=\e^{+\rmi(\alp{e}{}+\alp{e}{1})\,H}\,\op{R}^{21}(\p{}|\p{1})
\end{equation}
 We see that $\op{G}_{S}^{1+\,0-}$ contains a factor:
\begin{equation}
\e^{\rmi\,\alp{e}{}\,(z+H)+\,\rmi\,\alp{0}{0}\,z_0}\,,
\end{equation}
which is negligible far from the lower boundary if we have $K_e''H\gg 1$, since
\begin{equation}
\e^{\rmi\,\alp{e}{}\,(z+H)}\sim 0\,.
\end{equation}
Similarly, we show that far from the boundaries, the other contributions to  $\delta \op{G}^{11}_{S}$ are
negligible compared to $\op{G}^{\infty}_{1}$. Thus, we replace  in equation \eref{C11SVr} the term
$\op{G}_{S}^{11}$ by $\op{G}^{\infty}_{1}$ and the operator $\op{t}^{11}_{S,\rr_i}$ by $\op{t}^{11}_{\rr_i}$,
and we have
\begin{equation}
\fl<\op{C}^{11}_{SV,\rr_i}>_{V;\rr_i}=\op{t}^{11}_{\rr_i}+n\,\op{t}_{\rr_i}^{11}\cdot\op{G}_{1}^{\infty}\cdot\int_{V_1}\rmd^3\rr_j
[g(||\rr_j-\rr_i||)-1]\,<\op{C}^{11}_{SV,\rr_j}>_{V;\rr_j}\,\label{C11SVr-2}
\end{equation}
In doing this, we neglect all the boundary effects in the calculation of the effective permittivity $\ep_e$, and
equations obtained are the same used to calculate the effective permittivity in an infinite random
medium~\cite{Kong,Kong2001-3,Korringa}. In an infinite medium, we can use the Fourier transform to write the
equation \eref{C11SVr-2}, which is defined by
\begin{equation}
f(\kk|\kk_0)=\iintr{}{0}\exp(-\rmi\kk\cdot\rr+\rmi\kk_0\cdot\rr_0)\,f(\rr,\rr_0)\,.
\end{equation}
In the Fourier space, the translational invariance of the infinite medium implied  the following property of the
scattering operators $\op{t}^{11}_{\rr_i}$ (\ref{AppB}):
\begin{equation}
\op{t}^{11}_{\rr_i}(\kk|\kk_{0})=\exp(-\rmi\,(\kk-\kk_0)\cdot\rr_i)\,\op{t}^{11}_o(\kk|\kk_0)\,,\label{invt11tr}
\end{equation}
where $\op{t}^{11}_o(\kk|\kk_0)=\op{t}^{11}_{\rr_i=\vec{0}}(\kk|\kk_0)$ is the scattering operator for a
particle located at the origin of the coordinate (see \cite{Kong,Kong2001-3,Lax}). Using the property of
\eref{invt11tr} and equation \eref{C11SVr-2}, we show that $<\op{C}^{11}_{SV,\rr_i}(\kk|\kk_0)>_{V;\rr_i}$
verifies also a property similar to \eref{invt11tr}:
\begin{equation}
<\op{C}^{11}_{SV,\rr_i}(\kk|\kk_0)>_{V;\rr_i}
=\exp(-\rmi\,(\kk-\kk_0)\cdot\rr_i)\,\op{C}^{11}_{o}(\kk|\kk_0)\,,\label{invC11SV}
\end{equation}
where we have defined
$\op{C}^{11}_{o}(\kk|\kk_0)=<\op{C}^{11}_{SV,\rr_i=\vec{0}}(\kk|\kk_0)>_{V;\rr_i=\vec{0}}$. By using the
properties (\ref{invt11tr}, \ref{invC11SV}) in equations (\ref{C11SVr-2},\ref{eff-1}), we obtain:
\begin{eqnarray}
\fl\ep_e\,K_{vac}^2\,\op{I}=\ep_1\,K_{vac}^2\,\op{I}+n\,\op{C}^{11}_{o}(\kk_0|\kk_0)\,,\label{eff-2}
\\\fl \op{C}^{11}_{o}(\kk|\kk_0)=\op{t}^{11}_{o}(\kk|\kk_0)+n\,\intk{1}\,h(\kk-\kk_1)
\,\op{t}_{o}^{11}(\kk|\kk_1)\cdot\op{G}_{1}^{\infty}(\kk_1)\cdot\op{C}^{11}_{o}(\kk_1|\kk_0)\,,\label{eqC11-3}
\end{eqnarray}
where
\begin{eqnarray}
\op{t}^{11}_{o}=\op{v}^{11}_{o}+\op{v}^{11}_{o}\cdot\op{G}_{1}^{\infty}\cdot\op{t}^{11}_{o}\,,\\
\op{v}^{11}_{o}(\rr,\rr_0)=\delta(\rr-\rr_0)\,\op{v}^{11}_{o}(\rr)\,,\\
\op{v}^{11}_{o}(\rr)=K_{vac}^2\,(\tilde\ep_s-\ep_e)\,\Theta_s(\rr)\op{I}
\end{eqnarray}
and
\begin{eqnarray}
h(\kk-\kk_1)=\intr\,\exp(-\rmi\,(\kk-\kk_1)\cdot\rr)\,[g(||\rr||)-1]\,,\\
\op{G}_{1}^{\infty}(\kk)=\intr{}\exp(-\rmi\kk\cdot\rr)\,\op{G}_{1}^{\infty}(\rr)\,.\label{fourierG1}
\end{eqnarray}
\begin{figure}[htbp]
   \centering
   		\psfrag{z}{$z$}
      \psfrag{ri}{$\rr_i$}
      \psfrag{C11}{$<\op{C}^{11}_{SV,\rr_i}(\kk|\kk_0)>_V$}
      \psfrag{C11}{}
      \psfrag{H}{$H$}
      \psfrag{m0}{$\ep_0$}
      \psfrag{m1}{$\ep_e$}
      \psfrag{m2}{$\ep_2$}
      \psfrag{es}{$\ep_s$}
      \psfrag{k}{$\vec{k}$}
      \psfrag{k0}{$\vec{k}_0$}
      \psfrag{Correla}{\begin{minipage}{5cm}Area where the \\particles are correlated\end{minipage}}
         \epsfig{file=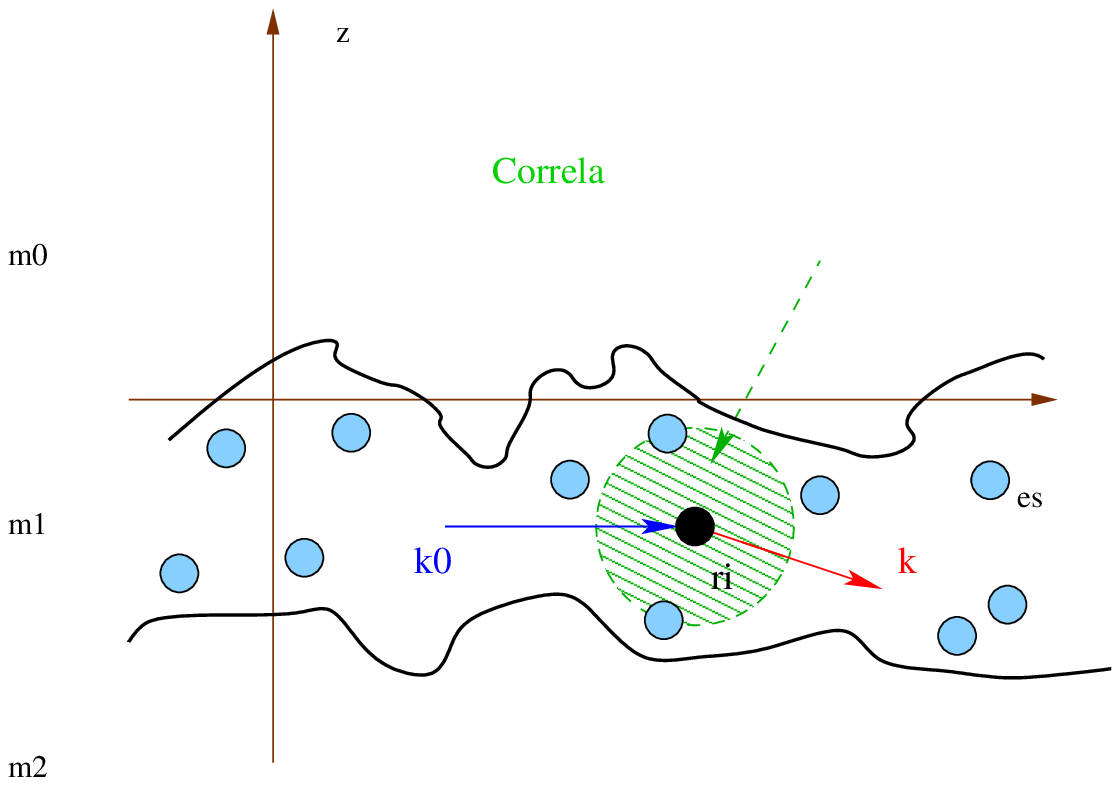,width=8cm}
         \caption{\label{Fig6bis}Graphical representation of 	 	 $<\op{C}^{11}_{SV,\rr_i}(\kk|\kk_0)>_V$}.
\end{figure}
In equation \eref{fourierG1}, we have used the translational invariance of the Green function:
$\op{G}_{1}^{\infty}(\rr,\rr_0)=\op{G}_{1}^{\infty}(\rr-\rr_0)$. Formula (\ref{eff-2}, \ref{eqC11-3}) is a
non-linear system of equations on the unknown $\ep_e(\omega)$. If we neglect the correlation between the
particles (i.e., $h(\vec{k}-\vec{k}_1)=0$) and define the Green function $\op{G}_1^{\infty}$ in replacing the
effective permittivity $\ep_e$ by $\ep_1$, we obtain the Foldy's approximation also called the independent
scattering approximation (ISA)~\cite{Sheng1,Ishi2,Lag1,Kong,Kong2001-3}:
\begin{equation}
\ep_e\,K_{vac}^2\,\op{I}=\ep_1\,K_{vac}^2\,\op{I}+n\,\op{t}^{11}_{o}(\kk_0|\kk_0)\Big|_{\ep_e\to\ep_1}\,.
\end{equation}
However, this result is greatly improved under the (CPA-QCA) approach since for Rayleigh scatterers, an
approximate formula for $\ep_e(\omega)$ can be derived from equations (\ref{eff-2}, \ref{eqC11-3}), which is a
generalization of the usual Maxwell-Garnett formula~\cite{Kong,Kong2001-3,Vries,Lag2,Kittel}. One can also
obtain an approximate formula for the effective permittivity, which at the same time contained the
Maxwell-Garnett formula and the Keller approximation~\cite{Soubret4}.
\section{Coherent field}
By using the expression in \eref{Chap3Eref} and the (CPA) condition \eref{PropCPA}, the average electric field
is
\begin{equation}
<\E_{SV}^{0}>_V=\E^{0i}+\E^{0s}_{S}\,.
\end{equation}
If we average over the surface disorder, we have
\begin{equation}
\ll\E_{SV}^{0}\gg_{SV}=\E^{0i}+<\E^{0s}_{S}>_S\,.
\end{equation}
Hence, for an incident plane wave,
\begin{equation}
\E^{0i}(\x,z)=\E^{0i}(\p{0})\,\e^{\rmi\p{0}\cdot\x-\rmi\alp{0}{0}\,z}\,,
\end{equation}
the field in the medium 0 is
\begin{eqnarray}
\fl\ll\E_{SV}^{0}(\x,z)\gg_{SV}=\E^{0i}(\p{0})\,\e^{\rmi\p{0}\cdot\x-\rmi\alp{0}{0}\,z}\,,\\
+\intp{}\e^{\rmi\p{}\cdot\x+\rmi\alp{0}{}z}\,<\op{S}^{0+0-}(\p{}|\p{0})>_S\cdot\E^{0i}(\p{0})\,,
\end{eqnarray}
where equations (\ref{S1-1-}, \ref{S1+1-}, \ref{S0+0-}) give us
\begin{eqnarray}
\RS{0+\,0-}=\op{R}^{10}+\op{T}^{01}\cdot\op{R}^{H\,21}\cdot[\op{I}^{1-1-}_{\perp}-\op{R}^{01}\cdot\op{R}^{H\,21}]^{-1}\op{T}^{10}\,.\label{expS}
\end{eqnarray}
For  statistical homogeneous rough surfaces, we have~\cite{Voro}:
\begin{equation}
<\op{S}^{0+0-}(\p{}|\p{0})>_S=(2\pi)^2\,\delta(\p{}-\p{0})\,\op{S}^{coh}(\p{0})\,,
\end{equation}
where $\op{S}^{coh}(\p{0})$ is a diagonal operator:
\begin{equation}
\op{S}^{coh}(\p{0})=S^{coh}(\p{0})_{VV}\,\,\evp{0+}{0}\evp{0-}{0}+S^{coh}(\p{0})_{HH}\,\,\ehp{0}\ehp{0}\,,
\end{equation}
or in a matrix form:
\begin{equation}
[\op{S}^{coh}(\p{0})]=\left(\begin{array}{cc}
S^{coh}(\p{0})_{VV}& 0\\
0 & S^{coh}(\p{0})_{HH}
\end{array}\right)\,,
\end{equation}
and
\begin{eqnarray}
\fl\ll\E_{SV}^{0}(\x,z)\gg_{SV}=\E^{0i}(\p{0})\,\e^{\rmi\p{0}\cdot\x-\rmi\alp{0}{0}\,z}+\op{S}^{coh}(\p{0})
\cdot\E^{0i}(\p{0})\,\e^{\rmi\p{0}\cdot\x+\rmi\alp{0}{0}z}\,.\no\\
\end{eqnarray}
Hence, the coherent field behaves as if the slab was an homogeneous medium of permittivity $\ep_e$ with  planar boundaries but with
modified Fresnel coefficients given by the two diagonal elements of $\op{S}^{coh}_{mat}(\p{0})$.
\section{Application}
Few approximate theories give explicit expression for the scattering operators \eref{expS} for a slab. Most of
them use the small-perturbation theory~\cite{Fuks1,Fuks2,soubret1,soubret2,Elson,Bousquet} to derive some
approximate expression of  $\RS{0+\,0-}$, but the perturbative development needs to go up to the second  order
to take into account the roughness of the surfaces. We must also mention that the Kirchhoff theory in the
geometrical optics limit and the full-wave method have been extended for a slab with rough
boundaries~\cite{Olidal,Bahar9}. However, we know that for the coherent part of the scattered field, the
exponential term present in the Kirchhoff theory gives an accurate description even in the small-perturbation
limit~\cite{Baylard}. Thus, under the Kirchhoff theory (or the first term of Small-Slope Approximation, which
gives the same results~\cite{Voro}), we have for Gaussian rough surfaces~\cite{Voro,Kong,Kong2001-3},
\begin{eqnarray}
<\op{R}^{10}(\p{}|\p{0})>_S=(2\pi)^2\,\delta(\p{}-\p{0})\,\op{r}^{10}(\p{0})\e^{-2\,\alp{0}{}^2\,\sigma_1^2}\,,\\
<\op{T}^{01}(\p{}|\p{0})>_S=(2\pi)^2\,\delta(\p{}-\p{0})\,\op{t}^{01}(\p{0})\e^{-(\alp{0}{0}-\alp{e}{0})^2\,\sigma_1^2/2}\,,\\
<\op{R}^{H\,21}(\p{}|\p{0})>_S=(2\pi)^2\,\delta(\p{}-\p{0})\,\op{r}^{H\,21}(\p{0})\e^{-2\alp{e}{0}^2\,\sigma_2^2}\,,\label{defrH21}\\
<\op{R}^{01}(\p{}|\p{0})>_S=(2\pi)^2\,\delta(\p{}-\p{0})\,\op{r}^{01}(\p{0})\e^{-2\,\alp{e}{0}^2\,\sigma_1^2}\,,\\
<\op{T}^{10}(\p{}|\p{0})>_S=(2\pi)^2\,\delta(\p{}-\p{0})\,\op{t}^{10}(\p{0})\e^{-(\alp{e}{0}-\alp{0}{0})^2\,\sigma_1^2/2}\,,\\
\end{eqnarray}
where $\sigma_1$, $\sigma_2$ are the rms-heights of the rough
surfaces:
\begin{equation}
\sigma_1=\sqrt{<h_1(\x)^2>_S}\,,\qquad \sigma_2=\sqrt{<h_2(\x)^2>_S}\,,
\end{equation}
and $\op{r}^{10}\,,\op{t}^{01}\,,\op{r}^{H\,21}\,,\op{r}^{01}\,,\op{r}^{01}$ are reflection operators for the
planar surface. We write these operators with matrices. For example,
\begin{equation}
\op{r}^{10}(\p{0})=r^{10}(\p{0})_{VV}\,\,\evp{0+}{0}\evp{0-}{0}+r^{10}(\p{0})_{HH}\,\,\ehp{0}\ehp{0}\end{equation} is written
\begin{equation}
[\op{r}^{10}(\p{0})]=\left(\begin{array}{cc}
r^{10}(\p{0})_{VV}& 0\\
0 & r^{10}(\p{0})_{HH}
\end{array}\right)\,,
\end{equation}
and we have:
\begin{eqnarray}
[\op{r}^{10}(\p{0})]=\left(\begin{array}{cc}
\case{\ep_e\alp{0}{0}-\ep_0\alp{e}{0}}{\ep_e\alp{0}{0}+\ep_0\alp{e}{0}} & 0\\
0 & \case{\alp{0}{0}-\alp{e}{0}}{\alp{0}{0}+\alp{e}{0}}
\end{array} \right)\,,\\
\left[\op{t}^{01}(\p{0})\right]=\left(\begin{array}{cc}
\case{2\,(\ep_0\,\ep_e)^{\frac{1}{2}}\,\alp{e}{0}}{\ep_e\alp{0}{0}+\ep_0\alp{e}{0}} & 0\\
0 & \case{2\,\alp{e}{0}}{\alp{0}{0}+\alp{e}{0}}
\end{array} \right)\,,\\
\left[\op{r}^{H\,21}(\p{0})\right]=\left(\begin{array}{cc}
\case{\ep_2\alp{e}{0}-\ep_e\alp{2}{0}}{\ep_2\alp{e}{0}+\ep_e\alp{2}{0}} & 0\\
0 & \case{\alp{e}{0}-\alp{2}{0}}{\alp{e}{0}+\alp{2}{0}}
\end{array} \right)\,\e^{2\rmi\alp{e}{0}H}\,,\label{Rh21MAt}\\
\left[\op{r}^{01}(\p{0})\right]=-[\op{r}^{10}(\p{0})]\,,\label{r10r01}\\
\left[\op{t}^{10}(\p{0})\right]=\left(\begin{array}{cc}
\case{2\,(\ep_0\,\ep_e)^{\frac{1}{2}}\,\alp{0}{0}}{\ep_e\alp{0}{0}+\ep_0\alp{e}{0}} & 0\\
0 & \case{2\,\alp{0}{0}}{\alp{0}{0}+\alp{e}{0}}
\end{array} \right)\,.
\end{eqnarray}
Hence, in  using the independent scattering approximation for the rough surfaces to calculate $<\RS{0+\,0-}>_S$,
\begin{eqnarray}
\fl<\RS{0+\,0-}>_S=<\op{R}^{10}>_S
+<\op{T}^{01}>_S\cdot<\op{R}^{H\,21}>_S\,\\
\cdot[\op{I}^{1-1-}_{\perp}-<\op{R}^{01}>_S\cdot<\op{R}^{H\,21}>_S]^{-1}\cdot<\op{T}^{10}>_S\,,
\end{eqnarray}
we obtain an approximate expression for the diagonal matrix $[\op{S}^{coh}]$ given by
\begin{eqnarray}
\fl[\op{S}^{coh}(\p{0})]=[\op{r}^{10}(\p{0})]\e^{-2\,\alp{0}{0}^2\,\sigma_1^2}+\e^{-(\alp{0}{0}-\alp{e}{0})^2\,\sigma_1^2-2\alp{e}{0}^2\,\sigma_2^2}\,[\op{t}^{01}(\p{0})]\cdot[\op{r}^{H\,21}(\p{0})]\no\\
\cdot\left([\op{I}_2]-\e^{-2\,\alp{e}{0}^2\,(\sigma_1^2+\sigma_2^2)}\,[\op{r}^{01}(\p{0})]\cdot[\op{r}^{H\,21}(\p{0})]\right)^{-1}\cdot[\op{t}^{10}(\p{0})]\,,\label{eqScoh}
\end{eqnarray}
where $[\op{I}_2]=diag(1,1)$ is the two dimensional identity matrix. Using the following identity
$[\op{r}^{10}(\p{0})]^2+[\op{t}^{01}(\p{0})]\cdot[\op{t}^{10}(\p{0})]=[\op{I}_2]$ ( which is the
conservation energy law for a planar surface) and the property in \eref{r10r01}, we rewrite  equation
\eref{eqScoh} under the following form:
\begin{eqnarray}
\fl[\op{S}^{coh}(\p{0})]=\Big([\op{c}^1(\p{0})]\cdot[\op{r}^{10}(\p{0})]+[\op{c}^2(\p{0})]\,\cdot[\op{r}^{H\,21}(\p{0})]\Big)\no\\
\cdot\Big([\op{I}_2]+[\op{c}^3(\p{0})]\cdot[\op{r}^{01}(\p{0})]\cdot[\op{r}^{H\,21}(\p{0})]\Big)^{-1}\,,\label{eqScoh-2}
\end{eqnarray}
with
\begin{eqnarray}
\fl[\op{c}^1(\p{0})]=\e^{-2\,\alp{0}{0}^2\,\sigma_1^2}\,[\op{I}_2]\,,\\
\fl[\op{c}^2(\p{0})]=\e^{-(\alp{0}{0}-\alp{e}{0})^2\,\sigma_1^2-2\alp{e}{0}^2\,
\sigma_2^2}\left([\op{I}_2]-(\e^{-(\alp{0}{0}+\alp{e}{0})^2\,\sigma_1^2}-1)[\op{r}^{10}(\p{0})]^2\right)\,,\no\\\\
\fl[\op{c}^3(\p{0})]=\e^{-2\,\alp{e}{0}^2\,(\sigma_1^2+\sigma_2^2)}\,[\op{I}_2]\,.
\end{eqnarray}
For a random medium with planar boundaries ($\sigma_1=\sigma_2=0$), we obtain
\begin{equation}
[\op{S}^{coh}(\p{0})]=\Big([\op{r}^{10}(\p{0})]+[\op{r}^{H\,21}(\p{0})]\Big)\cdot
\Big([\op{I}_2]+[\op{r}^{10}(\p{0})]\cdot[\op{r}^{H\,21}(\p{0})]\Big)^{-1}\,,\label{eqScoh-3}
\end{equation}
which is a diagonal matrix which contains the usual reflection coefficients for a planar slab separating three
homogeneous medium with the permittivities $\ep_0,\ep_e,\ep_2$. (See
references~\cite{Kong,Kong2001-3,Fuks1,Fuks2}.) In comparing  expression \eref{eqScoh-2} and \eref{eqScoh-3}, we
see that the rough surfaces modify the reflection coefficients for a planar slab by adding new factors
$[\op{c}^1]$,$[\op{c}^2]$,$[\op{c}^3]$. The random medium doesn't change the form of the reflection
coefficients but only the permittivity $\ep_1$ of the initial medium by an effective one $\ep_e$. Furthermore,
if the random medium is highly scattering and thick, the imaginary part of the effective permittivity $\ep_e$ is
important, and the factor $\exp(2\,\rmi\,\alp{e}{0}\,H)$ in the expression \eref{Rh21MAt} of $[\op{r}^{H\,21}]$ is
very small. The contribution  $[\op{r}^{H\,21}]$ in  equation \eref{eqScoh-3} becomes negligible compared to
$[\op{r}^{10}(\p{0})]$, and  we have
\begin{equation}
[\op{S}^{coh}(\p{0})]=\e^{-2\,\alp{0}{0}^2\,\sigma_1^2}\,[\op{r}^{10}(\p{0})]\,,
\end{equation}
which is the Kirchhoff term for a slab separating two
semi-infinite media with the permittivity $\ep_0$ and $\ep_e$. In
this case, the lower boundary doesn't contribute to the coherent
field.
\section{Conclusion}
We have considered the scattering of an electromagnetic wave by a random medium with rough boundaries. We have
formulated the solution of this problem using two kinds of Green functions. The first one describes the
scattering by the rough surfaces and the random medium, and the other  represents the scattering by an
homogeneous slab with rough boundaries. As equations obtained are similar to those used in scattering theory
by an infinite random medium, we were able to  introduce the coherent potential with the quasi-crystalline
approximation to calculate the effect of the random medium on the coherent field. With this approach, the random
medium contribution is taken into account by an effective medium permittivity. The surface scattering
contributions on the coherent field are included in the scattering operator of the system, which describes  the
scattering by the rough boundaries. This operator can be approximated using the usual scattering theories by
rough surface like the small-perturbation, the Kirchhoff, or other more sophisticated theories. To derive these
results, we have supposed that the slab is sufficiently thick to insure, for one hand,  that their exist a layer
($-H+max_{\x}h_2(\x)<z<min_{\x}h_1(\vec{x})$) between the two rough boundaries which contains the scatterers,
and in second hand,  that  the effective permittivity $\ep_e(\omega)$ doesn't not depend on the boundaries
($K_e''H\gg 1 $).

 In the following papers, we will use our Green function
formulation of the scattering problem to derive a radiative transfer equation describing the scattered
incoherent intensity. Furthermore, we will investigate the case of an highly scattering medium where a vectorial
diffusion approximation permits simplifying the radiative transfer equation.
\appendix
\section{Green functions}
\label{AppA}
\subsection{Scalar Green function}
The solution of
\begin{equation}
(\Delta+K_0^2)\,G_0(\rr,\rr_0)=-\dirac{\rr-\rr_0}\label{Gscalaire}
\end{equation}
in an infinite medium which satisfies the radiation condition at infinity is a generalized function given by
\begin{equation}
G^{\infty}_0(\rr-\rr_0)=P.V.\frac{\e^{\rmi\,K_0\,||\rr-\rr_0||}}{4\pi\,||\rr-\rr_0||}\,
,\label{PVspherique}\end{equation} where $P.V.$ is the principal value defined by
\begin{equation}
\fl P.V.\intr{0} G^{\infty}_0(\rr-\rr_0)\,\phi(\rr_0)\,=\lim_{a\rightarrow
0}\int_{V-V_a(\rr)}\;\;\rmd^3\rr_0\,G^{\infty}_0(\rr-\rr_0)\,\phi(\rr_0)\, ,\label{PVscal}
\end{equation}
where $\phi(\rr)$ is a test function, $V_a(\rr_0)$ is an exclusion volume with size $a$ around the singularity
located at $\rr$. In equation \eref{PVspherique}, the exclusion volume is a sphere~\cite{Hanson}. This
generalized function can be represented as the usual spherical function for $\rr\neq \rr_0$,
\begin{equation}
G^{\infty}_0(\rr-\rr_0)=\frac{\e^{\rmi\,K_0\,||\rr-\rr_0||}}{4\pi\,||\rr-\rr_0||}\,, ,\end{equation}

Using Fourier transform and the residue theorem, we can also write this function under the following form:
\begin{equation}
G^{\infty}_0(\rr)=\frac{\rmi}{2}\,\intp{} \e^{\rmi\,\p{}\cdot\x+\rmi \alp{0}{}\,|z|}\frac{1}{\alp{0}{}}\qquad
\rr\neq \vec{0}\,,\label{An3Weyl}
\end{equation}
The general solution of equation \eref{Gscalaire} can be expressed with equation \eref{An3Weyl} as a generalized
function:
\begin{equation}
G^{\infty}_0(\rr)=P.V.\frac{\rmi}{2}\,\intp{} \e^{\rmi\,\p{}\cdot\x+\rmi
\alp{0}{}\,|z|}\frac{1}{\alp{0}{}}\qquad \rr\neq \vec{0}\,,\label{An3Weyl2}
\end{equation}
where the exclusion volume used in the definition of the principal value \eref{PVscal} is a pillbox of arbitrary
cross section but thin in the z direction due to the term $|z|$ in the expression \eref{An3Weyl2} of the Green
function~\cite{Bladel,Yag}.
\subsection{Dyadic Green function}
The solution of
\begin{equation}
\nabla\times\nabla\times\op{G}_{0}^{\,\infty}(\rr,\rr_0)-K_{0}^2\,\op{G}^{\,\infty}_{0}(\rr,\rr_0)=
\delta(\rr-\rr_0)\,\op{I}\,
\end{equation}
in an infinite medium is a generalized function given by:
\begin{equation}
\op{G}_{0}^{\,\infty}(\rr-\rr_0)=\left(\op{I}+\frac{1}{K_{0}^2}\nabla\nabla\right)G_{0}^{\infty}(\rr-\rr_0)\,,
,\label{G0vect}
\end{equation}
which is short notation for
\begin{equation}
\fl\intr{0}
\op{G}_{0}^{\,\infty}(\rr-\rr_0)\,\phi(\rr_0)=\left(\op{I}+\frac{1}{K_{0}^2}\nabla\nabla\right)\intr{0}
G_{0}^{\infty}(\rr-\rr_0)\,\phi(\rr_0)\, ,\label{PVG0vect}
\end{equation}
where $\nabla\nabla f(\rr)=\nabla[\nabla f(\rr)]$. When we use the representation \eref{PVspherique} or
\eref{An3Weyl2} in \eref{G0vect}, the action of $\nabla\nabla$ on the exclusion volume $V_a(\rr)$ produces a
singularity:
\begin{equation}
\fl\op{G}_{0}^{\,\infty}(\rr-\rr_0)=P.V.\left(\op{I}+\frac{1}{K_{0}^2}
\nabla\nabla\right)G_{0}^{\infty}{\rr-\rr_0)}-\frac{1}{K_0^2}\delta(\rr-\rr_0)\,\op{L}\,, \label{decompoG0vect}
\end{equation}
where the principal value is defined by
\begin{eqnarray}
\fl P.V.\intr{0} \phi(\rr_0)\,\left(\op{I}+\frac{1}{K_{0}^2}\nabla\nabla\right)G_{0}^{\infty}(\rr-\rr_0)\,\no\\
=\lim_{a\rightarrow 0}\int_{V-V_a(\rr)}\;\;\rmd^3\rr_0\,\phi(\rr_0)\,
\left(\op{I}+\frac{1}{K_{0}^2}\nabla\nabla\right)G_{0}^{\infty}(\rr-\rr_0)\,.\label{PVscal-2}
\end{eqnarray}
The operator $\op{L}$ depends on the exclusion volume chosen; for a spherical volume, we have $\op{L}=\op{I}/3$,
and for a pillbox thin in the z direction, we have $\op{L}=\ez\,\ez$. As $\rr\neq \rr_0$ in the principal value
term, we can use the representation \eref{An3Weyl} to calculate the first term in \eref{decompoG0vect} and we
obtain
\begin{eqnarray}
\fl\op{G}_0^{\,\infty}(\rr,\rr_0)=\frac{\rmi}{2}\,P.V.\intp{0}\e^{\rmi\,\p{0}\cdot(\x-\x_0)+\rmi\,\alp{0}{0}\,|z-z_0|}\,(\op{I}-\hvec{k}^{0\,\pm}_{\p{0}}\hvec{k}^{0\,\pm}_{\p{0}})\,\frac{1}{\alp{0}{0}} \no\\
-\frac{1}{K^2_0}\delta(\rr-\rr_0)\op{L}\,, \label{An3G0VPWeyl}
\end{eqnarray}
where $\hvec{k}^{0\,\pm}_{\p{0}}=
\p{0}+sgn(z-z_0)\,\alp{0}{0}\,\ez$ and
\begin{equation}
sgn(z)=\left\{
\begin{array}{lc}
  +1 \, ,& \mbox{if}\quad z>0\\
  0 \, & \mbox{if}\quad z=0\\
  -1 \, & \mbox{if} \quad z< 0
\end{array} \right. \, .
\end{equation}
The upperscript sign in $\hvec{k}^{0\,\pm}$ is given by the sign
of the function $sgn(z-z_0)$.
\section{Transition operator for one scatterer}
\label{AppB} The electric field produced by an incident wave $\E^{i}(\rr)$ scattered by a spherical particle of
radius $r_s$, located at $\rr_j$, with a permittivity $\ep_a$, and surrounded by an infinite medium of
permittivity $\ep_b$ is given by~\cite{Kong,Kong2001-3,Lax}:
\begin{equation}
\vec{E}(\rr)=\E^{i}(\rr)+\int
\rmd^3\rr_1\,\,\op{G}^{\infty}_b(\rr,\rr_1)\cdot
\op{v}^1_{\rr_j}(\rr_1)\cdot\vec{E}(\rr_1)\,,\label{App2-0}
\end{equation}
where
\begin{equation}
\vec{v}^1_{\rr_j}(\rr_1)= K_{vac}^2\,(\ep_a-\ep_b)\,\Theta_s(\rr_1-\rr_j)\,\op{I}\,,\label{Chap3ParDefv1j}
\end{equation}
and
\begin{equation}
\Theta_s(\rr)=\left\{\begin{array}{cc}1 & \mbox{if}\quad ||\rr||<r_s\\
  0 & \mbox{if}\quad ||\rr||>r_s\end{array}\right.\,.\label{ChapDefThetad}
\end{equation}
The Green function $\op{G}^{\infty}_b(\rr,\rr_1)$ is defined by
\begin{eqnarray}
\op{G}_{b}^{\,\infty}(\rr-\rr_0)&=\left(\op{I}+\frac{1}{K_{b}^2}\nabla\nabla\right)G_{b}^{\infty}(\rr-\rr_0)\,,
\label{greenb}\\
G_{b}^{\infty}(\rr-\rr_0)&=P.V.\frac{\e^{\rmi\,K_b\,||\rr-\rr_0||}}{4\pi\,||\rr-\rr_0||}\,,
\end{eqnarray}
where $K_b^2=\ep_b\,K_{vac}^2$. The transition operator for one particle is defined by:
\begin{equation}
\vec{E}(\rr)=\E^{i}(\rr)+\int \rmd^3\rr_1\,\rmd^3\rr_2\,\,\op{G}^{\infty}_b(\rr,\rr_1)\cdot
\op{t}^{11}_{\rr_j}(\rr_1,\rr_2)\cdot\vec{E}^{i}(\rr_2)\,.\label{App2-1}
\end{equation}
In comparing the definition in \eref{App2-1} with equation \eref{App2-0}, we obtain
\begin{equation}
\op{t}^{11}_{\rr_j}(\rr,\rr_0) =\op{v}^1_{\rr_j}(\rr)\,\delta(\rr-\rr_0)+\int
\rmd^3\rr_1\,\,\op{v}^1_{\rr_j}(\rr)\cdot\op{G}^{\infty}_b(\rr,\rr_1)\cdot
\op{t}^{11}_{\rr_j}(\rr_1,\rr_0)\,.\label{Chap3E1transtv}
\end{equation}
In the Fourier space, we have
\begin{equation}
\op{t}^{11}_{\rr_j}(\kk|\kk_0)
=\op{v}^1_{\rr_j}(\kk-\kk_0)+\intk{1}\,\,\op{v}^1_{\rr_j}(\kk-\kk_1)\cdot\op{G}^{\infty}_b(\kk_1)\cdot
\op{t}^{11}_{\rr_j}(\kk_1|\kk_0)\,,\label{t11-one}
\end{equation}
with \begin{equation} \op{v}^1_{\rr_j}(\kk-\kk_0)=\int \rmd^3 \rr\,
\e^{-\rmi(\kk-\kk_0)\cdot\rr}\,\op{v}^1_{\rr_j}(\rr)\, ,
\end{equation}
and
\begin{equation}
\op{G}^{\infty}_{b}(\kk_1)=\int \rmd^3 \rr\, \e^{-\rmi\kk_1\cdot\rr}\,\op{G}^{\infty}_b(\rr)\, .
\end{equation}
We easily check that $\op{v}^1_{\rr_j}$ verifies the following property: \begin{equation}
\op{v}^1_{\rr_j}(\kk-\kk_0)=\e^{-\rmi\,(\kk-\kk_0).\rr_j}\,\op{v}^1_o(\kk-\kk_0)\,,\label{Chap3vjvo}
\end{equation}
where $\op{v}^1_{o}(\kk-\kk_0)=\op{v}^1_{\rr_j=\vec{0}}(\kk-\kk_0)$. In iterating equation \eref{t11-one}, and
using the property \eref{Chap3vjvo}, we demonstrate that
\begin{equation}
  \op{t}^{11}_{\rr_j}(\kk|\kk_0)=\exp(-\rmi\,(\kk-\kk_0)\cdot\rr_j)\,\op{t}^{11}_{o}(\kk|\kk_0)\,,\label{fromdemo}
 \end{equation}
where $\op{t}^{11}_{o}=\op{t}^{11}_{\rr_j=\vec{0}}$ is the transition operator for a particle located at the
origin of the coordinate. If we consider an incident plane wave $\E^{i}(\rr)$,
\begin{equation}
\E^{i}(\rr)=\E^{i}(\vec{k}_0)\,\e^{\rmi\vec{k}_0\cdot\rr}\,,
\end{equation}
transverse to the propagation direction $\vec{k}_0$:
\begin{equation}
\hvec{k}_0\cdot\E^{i}(\vec{k}_{0})=0\quad\Longleftrightarrow \quad
(\op{I}-\hvec{k}_0\hvec{k}_0)\cdot\E^{i}(\vec{k}_{0})=\E^{i}(\vec{k}_{0})
\end{equation}
 where $\vec{k}_0=K_b\,\hvec{k}_0$, and
$\hvec{k}_0\cdot\hvec{k}_0=1$. The far-field scattered by a particle located at the origin is given by
\begin{equation}
\fl K_b||\rr||\gg 1\,, \qquad
\vec{E}(\rr)=\E^{i}(\rr)+\frac{\e^{\rmi\,K_b\,||\rr||}}{4\,\pi\,||\rr||}\,(\op{I}-\hvec{k}\hvec{k})
\cdot\op{t}^{11}_o(\kk|\kk_0)\cdot(\op{I}-\hvec{k}_0\hvec{k}_0)\cdot\E^{i}(\kk_{0})\,,\label{Chap3deftik}
\end{equation}
where
\begin{equation}
\op{t}^{11}_o(\kk|\kk_0)=\iintr{}{0} \e^{-\rmi \kk\cdot
  \rr+\rmi\,\kk_0\cdot \rr_0} \,\op{t}^{11}_o(\rr|\rr_0)\,.\label{Chap3eqItjvj}
\end{equation}
We have used the following far-field approximation to derive the
equation \eref{Chap3deftik}:
\begin{equation}
\op{G}_b^{\,\infty}(\rr,\rr_1)\approx\frac{\e^{\rmi\,K_b\,||\rr||}}{4\,\pi\,||\rr||}
(\op{I}-\hvec{k}\hvec{k})\,\e^{-\rmi\kk\cdot\rr_1}\,,\label{Chap3ApproxG0vect}
\end{equation}
where $\kk=K_b\,\hvec{r}$. Usually, the far field scattered by a
particle is written in the following
form~\cite{Ishi2,Hulst1,Bohren}:
\begin{equation}
K_b||\rr||\gg 1\,, \qquad
\vec{E}^s(\rr)=\frac{\e^{\rmi\,K_b\,||\rr||}}{||\rr||}\,\op{f}(\hvec{k}|\hvec{k}_0)\cdot\E^{i}(\kk_{0})\,.\label{Chap3deftik-2}
\end{equation}
For spherical scatterer, an exact expression of this operator is well-known and given by the Mie
theory~\cite{Ishi2,Hulst1,Kerker,Bohren}. In comparing equations \eref{Chap3deftik} and \eref{Chap3deftik-2}, we
have the following relationship between $\op{f}(\hvec{k}|\hvec{k}_0)$ and $\op{t}^{11}_o(\kk|\kk_0)$ :
\begin{equation}
4\pi\,\op{f}(\hvec{k}|\hvec{k}_0)=(\op{I}-\hvec{k}\hvec{k})\cdot\op{t}^{11}_o(\kk|\kk_0)\cdot(\op{I}-\hvec{k}_0\hvec{k}_0)\,,\label{Chap3lientjf}
\end{equation}
where $\kk=K_b\,\hvec{k}$, and $\kk_0=K_b\,\hvec{k}_0$. The operator $\op{t}^{11}_o(\kk|\kk_0)$ is a
generalization of the scattering amplitude $\op{f}(\hvec{k}|\hvec{k}_0)$ since it contains also the near-field
component scattered by the particle. Furthermore, we see  that the far-field component is obtained in taking
only the transversal components of $\op{t}^{11}_o(\kk|\kk_0)$ (due to the projectors $\op{I}-\vec{k}\vec{k}$ and
$\op{I}-\vec{k}_0\vec{k}_0$) and using an on-shell approximation (since
$\vec{k}\cdot\vec{k}=\vec{k}_0\cdot\vec{k}_0=K_b$).
\section{Reciprocity of $\op{C}^{11}_{o}(\vec{k}|\vec{k}_0)$}
In the section \ref{secCPA}, we have used the following decomposition of the operator $\op{T}^{11}_{SV}$:
\begin{equation}
\op{T}^{11}_{SV}=\op{V}^{11}+\op{V}^{11}\cdot\op{G}_{S}^{11}\cdot\op{T}^{11}_{SV}\,.\label{defTg}
\end{equation}
But we could have used this equivalent formulation:
\begin{equation}
\op{T}^{11}_{SV}=\op{V}^{11}+\op{T}_{SV}^{11}\cdot\op{G}_{S}^{11}\cdot\op{V}^{11}\,.\label{defTd}
\end{equation}
According to the section \ref{secCPA}, we define a new  operator $\optilde{V}^{11}$ 
such that:
\begin{eqnarray}
\optilde{V}^{11}=\op{V}^{11}+\op{W}^{11}\,,
\end{eqnarray}
where $\op{W}^{11}$ is defined by equation \eref{defW11}.
The operator $\op{T}^{11}_{SV}$ can also be decomposed in defining new operators $\optilde{T}^{11}_{SV}$ or  $\optilde{TT}^{11}_{SV}$ and using either equation \eref{defTg}
or \eref{defTd}:
\begin{eqnarray}
\fl\mbox{with the Eq. \eref{defTg}}\quad \left\{ \begin{array}{l}
\optilde{T}^{11}_{SV}=\op{T}^{11}_{SV}+\op{Q}^{11}_{SV}\,,\\
\op{Q}^{11}_{SV}=\op{W}^{11}+\op{W}^{11}\cdot\op{G}_{S}^{11}\cdot\op{T}^{11}_{SV}\,,
\end{array}\right.\label{AppC-g}\\
\fl\mbox{with the Eq. \eref{defTd}}\quad \left\{ \begin{array}{l}
\optilde{TT}^{11}_{SV}=\op{T}^{11}_{SV}+\op{QQ}^{11}_{SV}\,,\\
\op{QQ}^{11}_{SV}=\op{W}^{11}+\op{T}^{11}_{SV}\cdot\op{G}_{S}^{11}\cdot\op{W}^{11}\,,
\end{array}\right.\label{AppC-d}
\end{eqnarray}
Following the demonstration of section \ref{secCPA}, we derive the following equations:
\begin{eqnarray}
\fl\left\{ \begin{array}{l}
\optilde{T}^{11}_{SV}=\sum_{i=1}^N \op{C}^{11}_{SV,\rr_i}\,,\label{AppC-1}\\
\op{C}^{11}_{SV,\rr_i}=\op{t}^{11}_{S,\rr_i}+\op{t}^{11}_{S,\rr_i}\cdot\op{G}_{S}^{11}\cdot\left(\sum_{j=1, j\neq i}^N \op{C}^{11}_{SV,\rr_j}-\op{Q}^{11}_{SV}\right)\label{AppC-2}
\end{array}\right.\\
\fl\left\{ \begin{array}{l}
\optilde{T}^{11}_{SV}=\sum_{i=1}^N \op{CC}^{11}_{SV,\rr_i}\,,\label{AppC-3}\\
\op{CC}^{11}_{SV,\rr_i}=\op{t}^{11}_{S,\rr_i}+\left(\sum_{j=1, j\neq i}^N \op{CC}^{11}_{SV,\rr_j}-\op{QQ}^{11}_{SV}\right)\cdot\op{G}_{S}^{11}\cdot\op{t}^{11}_{S,\rr_i}
\label{AppC-4}
\end{array}\right.
\end{eqnarray}
If we define the average of the operators  $\op{C}^{11}_{SV,\rr_i}$ and $\op{CC}^{11}_{SV,\rr_i}$ at the origin ($\rr_i=\vec{0}$) by:
\begin{eqnarray}
\op{C}^{11}_{o}=<\op{C}^{11}_{SV,\rr_i=\vec{o}}>_{V;\rr_i=\vec{o}}\,,\\
\op{CC}^{11}_{o}=<\op{CC}^{11}_{SV,\rr_i=\vec{o}}>_{V;\rr_i=\vec{o}}\,,\\
\end{eqnarray}
we obtain, by using equations (\ref{AppC-1}-\ref{AppC-4}), the following expressions:
\begin{eqnarray}
\fl \op{C}^{11}_{o}(\kk|\kk_0)=\op{t}^{11}_{o}(\kk|\kk_0)+n\,\intk{1}\,h(\kk-\kk_1)
\,\op{t}_{o}^{11}(\kk|\kk_1)\cdot\op{G}_{1}^{\infty}(\kk_1)\cdot\op{C}^{11}_{o}(\kk_1|\kk_0)\,,\nonumber\\\label{AppC-Four1}\\
\fl \op{CC}^{11}_{o}(\kk|\kk_0)=\op{t}^{11}_{o}(\kk|\kk_0)+n\,\intk{1}\,h(\kk_1-\kk_0)
\,\op{CC}^{11}_{o}(\kk|\kk_1)\cdot\op{G}_{1}^{\infty}(\kk_1)\cdot\op{t}_{o}^{11}(\kk_1|\kk_0)\,,\nonumber\\\label{AppC-Four2}
\end{eqnarray}
Furthermore, under the (CPA) approximation we have $<\op{T}^{11}_{SV}>_V=0$, and from equations (\ref{AppC-g},\ref{AppC-d}), we deduce that $<\op{Q}^{11}_{SV}>_V=<\op{QQ}^{11}_{SV}>_V=\op{W}^{11}$
and then 
\begin{equation}
<\optilde{T}^{11}_{SV}>_V=<\optilde{TT}^{11}_{SV}>_V\,.\label{eqTTT}
\end{equation}
By using the decomposition (\ref{AppC-1}, \ref{AppC-3}) and the definition of the conditional average $<\,\, >_{V;\rr_i}$,  equation \eref{eqTTT} can be written as:
\begin{equation}
n\,\int_{V_1}\rmd^3\rr_i\,<\op{C}^{11}_{SV,\rr_i}>_{V;\rr_i}=n\,\int_{V_1}\rmd^3\rr_i\,<\op{CC}^{11}_{SV,\rr_i}>_{V;\rr_i}\,.
\end{equation}
This identity is valid whatever the volume $V_1$ and the position $\rr_i$ of the scatterer, and thus we have:
\begin{equation}
\op{C}^{11}_{o}(\kk|\kk_0)=\op{CC}^{11}_{o}(\kk|\kk_0)\,.\label{eqCCC}
\end{equation}
Hence, the operator $\op{C}^{11}_{o}(\kk|\kk_0)$ satisfies the following equations:
\begin{eqnarray}
\fl \op{C}^{11}_{o}(\kk|\kk_0)=\op{t}^{11}_{o}(\kk|\kk_0)+n\,\intk{1}\,h(\kk-\kk_1)
\,\op{t}_{o}^{11}(\kk|\kk_1)\cdot\op{G}_{1}^{\infty}(\kk_1)\cdot\op{C}^{11}_{o}(\kk_1|\kk_0)\,,\nonumber\\
\fl \op{C}^{11}_{o}(\kk|\kk_0)=\op{t}^{11}_{o}(\kk|\kk_0)+n\,\intk{1}\,h(\kk_1-\kk_0)
\,\op{C}^{11}_{o}(\kk|\kk_1)\cdot\op{G}_{1}^{\infty}(\kk_1)\cdot\op{t}_{o}^{11}(\kk_1|\kk_0)\,.\nonumber
\end{eqnarray}
However, since the operator $\op{t}^{11}_{o}$ is reciprocal and $\op{G}_{1}^{\infty}(\kk_1)=\op{G}_{1}^{\infty}(-\kk_1)$, we easily show   using equations (\ref{AppC-Four1},\ref{AppC-Four2}) that $\op{C}^{11}_{o}(\kk|\kk_0)=[\op{CC}^{11}_{o}(-\kk_0|-\kk)]^{T}$ where $^T$ is the transpose of the operator. From the identity \eref{eqCCC}, we conclude that the operator $\op{C}^{11}_{o}(\kk|\kk_0)$ is reciprocal:
\begin{equation}
\op{C}^{11}_{o}(\kk|\kk_0)=[\op{C}^{11}_{o}(-\kk_0|-\kk)]^{T}\,.
\end{equation}
\section*{References}
\bibliographystyle{Wavesunsrt}
\bibliography{Livres_Bib,Articles_Bib}
\end{document}